\documentclass[aps,prb,floatfix,superscriptaddress,noshowpacs,twocolumn]{revtex4-2}
\usepackage{subcaption}
\usepackage{graphicx,amsmath,amssymb,epstopdf,wasysym}
\usepackage[space]{grffile}
\usepackage{hyperref}
\usepackage{bm}
\usepackage[dvipsnames]{xcolor}
\usepackage{braket,amsfonts}
\usepackage{dsfont}
\usepackage{float} 
\usepackage{placeins}
\allowdisplaybreaks
\newcommand{\ca}{c^{\phantom{\dagger}}}
\newcommand{\cc}{c^\dagger}

\newcommand{\be}{\begin{equation}}
\newcommand{\ee}{\end{equation}}
\newcommand{\bga}{\begin{gather}}
\newcommand{\ega}{\end{gather}}
\newcommand{\bea}{\begin{eqnarray}}
\newcommand{\eea}{\end{eqnarray}}
\newcommand{\dagga}{{\phantom{\dagger}}}
\newcommand{\bR}{\mathbf{R}}

\newcommand{\bM}{\mathbf{M}}

\newcommand{\bk}{\mathbf{k}}

\newcommand{\bRp}{\mathbf{R'}}
\newcommand{\bx}{\mathbf{x}}
\newcommand{\by}{\mathbf{y}}
\newcommand{\bz}{\mathbf{z}}

\newcommand{\Ima}{\text{Im}}
\newcommand{\Rea}{\text{Re}}
\newcommand{\dis}{\displaystyle}

\newcommand{\up}{\uparrow}
\newcommand{\down}{\downarrow}
\newcommand{\fract}[2]{\frac{\dis \;#1\;}{\dis \;#2\;}}

\newcommand{\eqn}[1]{(\ref{#1})}

\newcommand{\ep}{{\epsilon}}

\newcommand{\bw}{\begin{widetext}}
\newcommand{\ew}{\end{widetext}}
\newenvironment{eqs}%
{\begin{equation} \begin{aligned}}%
{\end{aligned} \end{equation} }
\newcommand{\beal}{\begin{eqs}}
\newcommand{\eal}{\end{eqs}}
\newcommand{\bd}[1]{{\boldsymbol{#1}}}
\newcommand{\esp}[1]{\text{e}^{#1}}
\newcommand{\bealn}{\beal\nonumber}

\begin{document}
\title{Signatures of the Fermi surface reconstruction of a doped Mott insulator in a slab geometry  
}
\author{Gregorio Staffieri}
\email{Contact author: gstaffie@sissa.it}
\affiliation{International School for Advanced Studies (SISSA), Via Bonomea 265, I-34136 Trieste, Italy} 
\author{Michele Fabrizio}
\affiliation{International School for Advanced Studies (SISSA), Via Bonomea 265, I-34136 Trieste, Italy} 

\begin{abstract} 
We investigate a hole-doped Mott insulator in a slab geometry using the dynamical cluster approximation. We show that the enhancement of the correlation strength at the surface results in the remarkable evolution of the layer-projected Fermi surface, which exhibits hole-like pockets in the superficial layers, but gradually evolves into a single electron-like surface in the innermost layers.
We further analyze the behavior of the Friedel oscillations induced by the surface and identify distinct signatures of the Fermi surface reconstruction as function of hole-doping. In addition, we introduce a computationally tractable quantity that diagnoses the same Fermi surface variation by the concurrent breakdown of Luttinger's theorem. Both the latter quantity and the Friedel oscillations serve as reliable indicators of the change in Fermi surface topology, without the need for any periodization in momentum space. 
\end{abstract}

\maketitle 

\section*{Introduction}
One of the many captivating features of high-T$_c$ copper-oxide superconductors is the evidence of 
a Fermi surface reconstruction upon hole doping \cite{Taillefer_2009,Suchitra-2015,Taillefer-ARCM2019}. In the overdoped regime, the Fermi surface is electron-like, large \cite{Hussey-Nature2008} and aligns with Luttinger's theorem \cite{Luttinger}. Conversely, in the opposite underdoped regime the Fermi surface consists of Fermi pockets \cite{Norman-Nature1998,Taillefer-Nature2007,
Taillefer-Nature2007-bis,Zhou-Nature2009,Harrison-NatPhy2020,Kunisada833} that appear to violate
Luttinger's theorem and coexist with a pseudogap at the antinodal points \cite{Timusk_1999,Shen-RMP2003,Shen-RMP2021}. In angle-resolved photoemission, these pockets resembles Fermi arcs \cite{Norman-Nature1998,Zhou-Nature2009}, which is likely due to the interference with a nearby Luttinger surface \cite{Rice-PRB2006,Rice-RPP2011,Alexei-RPP2019,Michele2}, 
the location in momentum space of the zeros of the Green's function at zero frequency \cite{Igor-PRB2003}, as opposed to the location of poles that defines the Fermi surface. 
Indeed, the existence of a Luttinger surface in the underdoped regime of cuprates explains both 
the presence of a pseudogap in the spectrum and the violation of Luttinger's theorem 
\cite{Heath_2020,Jan-PRB2022}. \\
Evidence of a similar doping-driven Fermi surface reconstruction has been also discovered in the simple single-band Hubbard model by means of various numerical techniques, ranging from variational \cite{Federico-PRB2012} and diagrammatic \cite{Georges-PRR2024} Monte Carlo methods, to cluster extensions of dynamical mean-field theory \cite{PhysRevB.74.125110,RevModPhys.77.1027,cluster2,PhysRevLett.102.056404,PhysRevB.82.134505,Sordi-PRL2010,PhysRevB.82.155101,PhysRevB.83.214522,PhysRevLett.110.216405,Georges-PNAS2018,Georges-PRX2018,PhysRevLett.120.067002,Imada-NatComm2023}. \\
In this study, we explore the intriguing effects of such Fermi surface reconstruction on doping a slab of a model Mott insulator. We investigate this phenomenon using the dynamical cluster approximation \cite{RevModPhys.77.1027}. The slab geometry presents several intriguing avenues that merit exploration. In particular, in Sec.~\ref{LT} we study  
the doped Mott insulating slab with open boundary conditions. We demonstrate that the effective enhancement of correlation strength at the surface, which is responsible for the phenomenon known as the surface dead layer effect \cite{dead-layer,Antonio-dead-layer}, results in a non-uniform distribution of doping that decreases as we approach the surface. Consequently, a layer-dependent Fermi surface reconstruction occurs that could be experimentally investigated in a similar manner to the dead layer effect \cite{Marino-dead-layer}. 
In Section~\ref{Friedel-oscillations}, we study the Friedel oscillations that arise from the open geometry of the slab. We show that these oscillations can serve as an additional tool to discern the changes in the size and shape of the Fermi surface as the doping level is increased. 
Finally, Section~\ref{conc} is devoted to Conclusions.

\section{Layer dependent Fermi surface reconstruction} \label{LT}
\begin{figure}[ht]
    \centering
    \includegraphics[width=0.46\textwidth]{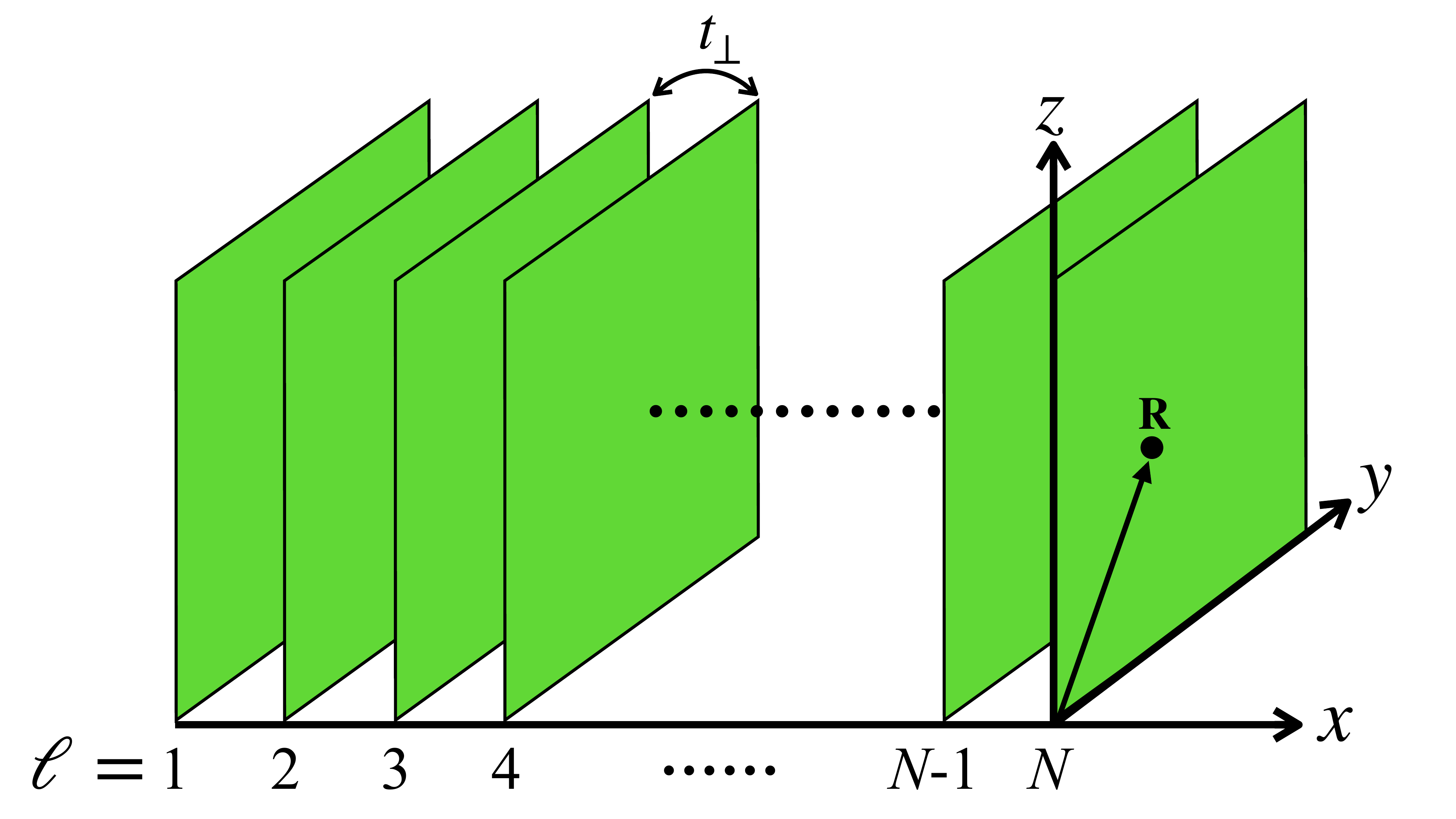}
            \caption{Sketch of the slab geometry for N layers, which are stacked along $\bx$ axis. The intra-layer sites are identified by the $\bR$ vector coordinates.}
    \label{fig:illustrazione_slab}
 \end{figure}

We consider a slab composed of $N$ layers, each of which is described by a two-dimensional (2D) Hubbard model with nearest neighbor hopping $t$. Nearest neighbor layers are in turn coupled to each other by an interlayer hopping $t_\perp$, as illustrated in Fig.~\ref{fig:illustrazione_slab}. The system has periodic boundary conditions along $\by$ and $\bz$, while open boundary conditions are imposed along $\bx$. The Hamiltonian reads 
 \beal
H&=-t\,\sum_{\ell=1}^{N}\,  \sum_{\langle \bR \bRp \rangle \sigma}\, \big( \cc_{\bR \ell \sigma }\ca_{\bRp \ell \sigma}+H.c. \big)  \\
&\qquad - \mu\,\sum_{\ell=1}^{N} \,\sum_{\bR \sigma} \,    n_{\bR \ell \sigma} + U\,\sum_{\ell=1}^{N} \,\sum_{\bR}\, 
    n_{\bR \ell \up} \,  n_{\bR \ell \down} \\
&\qquad  -t_{\perp}\, \sum_{\ell=1}^{N-1}\, \sum_{\bR \sigma }\, \big( \cc_{\bR \ell \sigma}\ca_{\bR \ell+1 \sigma}+H.c. \big)\,.
\label{Ham}
\eal
The operators $c^\dagga_{\bR\ell\sigma}$ and $c^\dagger_{\bR\ell\sigma}$ annihilate and create, respectively, one electron on site $\bR$ of  
layer $\ell=1...N$ with spin $\sigma=\up,\down$, while $n_{\bR\ell\sigma} = c^\dagger_{\bR\ell\sigma}\,c^\dagga_{\bR\ell\sigma}$. The other parameters are $U$, the on-site Hubbard repulsion, and $\mu$, the chemical potential. \\ 
We study the Hamiltonian \eqn{Ham} by means of dynamical cluster approximation (DCA) \cite{RevModPhys.77.1027}. Within DCA, the Brillouin zone is partitioned into patches and the self-energy assumed constant within each patch, resulting in a piecewise function of momentum. We use the quantum impurity solver implemented within the TRIQS library \cite{PARCOLLET2015398}, which employs a continuous time quantum Monte Carlo (CTQMC) evaluation of the partition function expanded in powers of the hybridization \cite{SETH2016274}. 
Hereafter, we consider a partition of the Brillouin zone in four patches centered at the high symmetry points, $\mathbf{\Gamma}=(0,0)$, $\mathbf{X}=(\pi,0)$, $\mathbf{Y}=(0,\pi)$ and $\bM=(\pi,\pi)$, with the same tiling of Ref.~\cite{DCA_first_paper}. This choice is the simplest one to access the antinodal points $\mathbf{X}$ and $\mathbf{Y}$, which are important to study the Fermi surface reconstruction \cite{PhysRevLett.120.067002}. \\ 
Layered systems, such as slabs, interfaces and heterostructures, have been extensively investigated by  DMFT \cite{Kondoprox,heterostructures,linearizedDMFT}, where the self-energy is assumed to be local in each layer but layer dependent, $ \Sigma_{\ell\ell'}(i \omega_{n})=\Sigma_{\ell}(i \omega_{n}) \delta_{\ell\ell'} $, where $\omega_n$ are Matsubara frequencies. We improve this description by including the coarse-grained momentum effects of DCA, thus a self-energy matrix with components 
\bealn 
\Sigma_{\ell \ell'}(i \omega_{n},\mathbf{P})=\delta_{\ell \ell'}\, \Sigma_{\ell}(i \omega_{n},\mathbf{P})\,,
\eal
where $\mathbf{P}$ is one of the four patches. 
Within this approximation, the lattice Green function $\hat{G}(i\omega_n,\bk)$ is a matrix in layer indices \cite{Kondoprox} whose inverse satisfies  
\be
\hat{G}(i \omega_n,\bk)^{-1}= i \omega_n+\mu-\ep(\bk)-
 \hat{\Sigma}(i \omega_n,\mathbf{P_{\bk}}) + \hat{t}_{\perp}\,,
\label{eqn:glat}
\ee   
where $\hat{t}_{\perp}$ is the matrix with elements $t_\perp$ just between nearest neighbor layers, 
$\bk$ the momentum in the $y$-$z$ plane, $\ep(\bk)=-2t(\cos k_y +\cos k_z )$, and $\mathbf{P_{\bk}}$  the patch reference momentum that contains $\mathbf{k}$. The local layer-dependent Green function is  obtained averaging the diagonal elements of $\hat{G}(i\omega_n,\bk)$ over each patch
\beal
G(i \omega_n, \mathbf{P},\ell)=\fract{1}{N_{\mathbf{P}}} \sum_{\bk \in \mathbf{P}}\,
 G_{\ell\ell}(i \omega_n,\bk)\,,
\label{eqn:gloc}
\eal
with $N_{\mathbf{P}}$ the number of $\bk$-points in the patch $\mathbf{P}$.
The patch-dependent dynamical Weiss field  $\mathcal{G}_0(i \omega_n, \mathbf{P}, \ell)$ at layer $\ell$ is computed through Dyson's equation
\be
\mathcal{G}_0(i \omega_n, \mathbf{P}, \ell)^{-1} = G(i \omega_n, \mathbf{P}, \ell)^{-1} + \Sigma_{\ell}(i \omega_n, \mathbf{P})\,,\nonumber
\ee
where $\Sigma_{\ell}(i \omega_n, \mathbf{P})$, 
$\mathbf{P}=\bd{\Gamma},\mathbf{X},\mathbf{Y},\mathbf{M}$, are the self-energies of four Anderson impurities 
coupled together by interaction and each hybridized to a conduction bath. 
The Weiss field is in turn used to compute the layer-$\ell$ partition function, thus closing the self-consistent loop. For simplicity, in what follows we take $ t_{\perp}=t$ the unit of energy.

\begin{figure}[h!]
    \centerline{
    \includegraphics[width=0.5\textwidth]{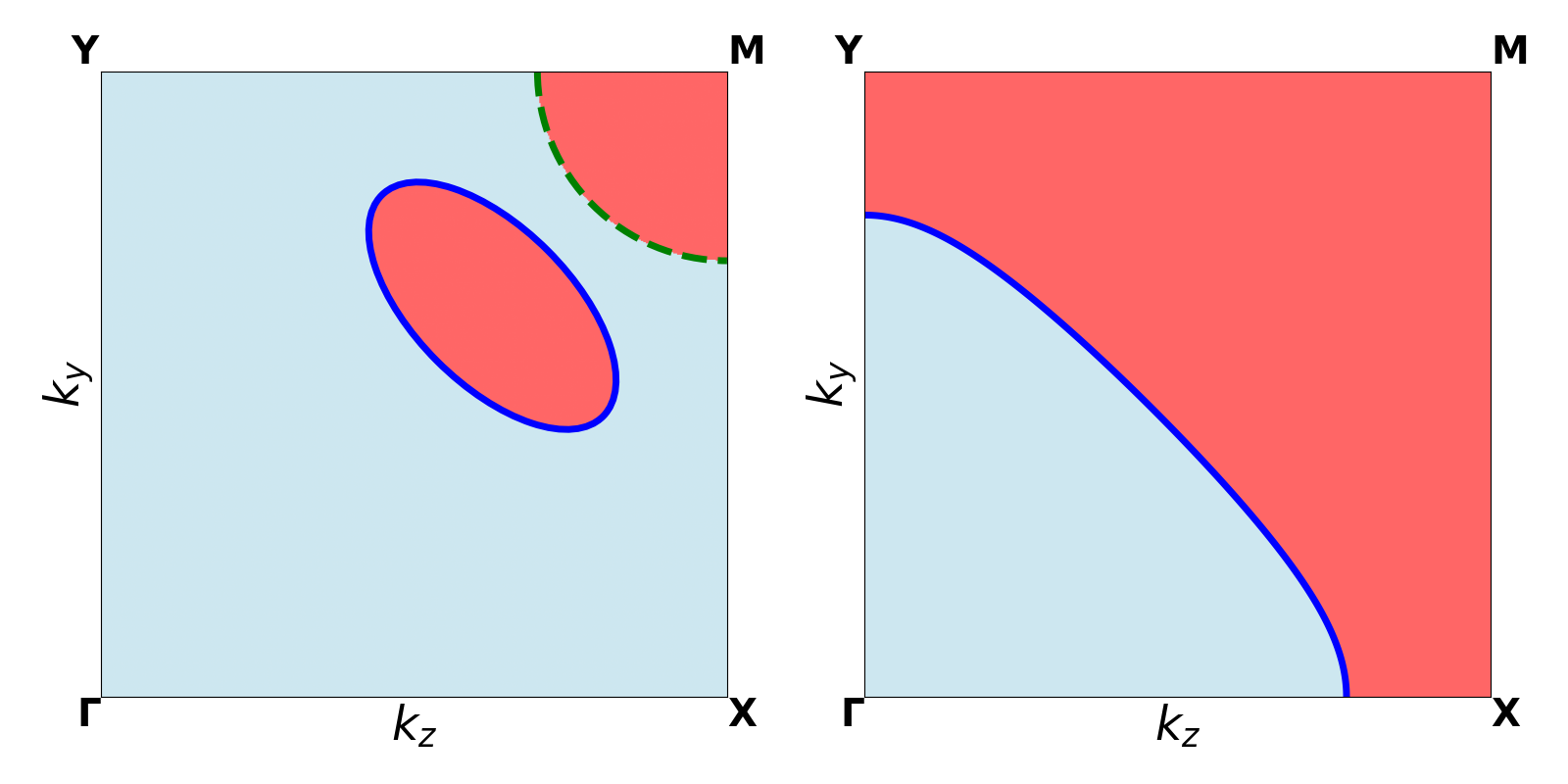}}
            \caption{ Left panel: sketch of the renormalized dispersion $r(\bk)=\ep(\bk)-\mu+\Rea \Sigma(0,\bk)$ in the upper right quarter of the Brillouin zone for the scenario proposed by  Ref.~\cite{PhysRevB.74.125110} in the case of an hole-like Fermi surface with Fermi pockets. The red (lightblue) color indicates $r(\bk)>0$ ($r(\bk)<0$). The blue solid line represents the Fermi surface, while the dashed green line indicates the Luttinger surface. Right Panel: same as left panel but for an electron-like Fermi surface that satisfies Luttinger theorem. }
    \label{fig:hole_vs_electron}
\end{figure}
\noindent

\begin{figure*}[ht]
    \centerline{\includegraphics[width=0.45\textwidth]{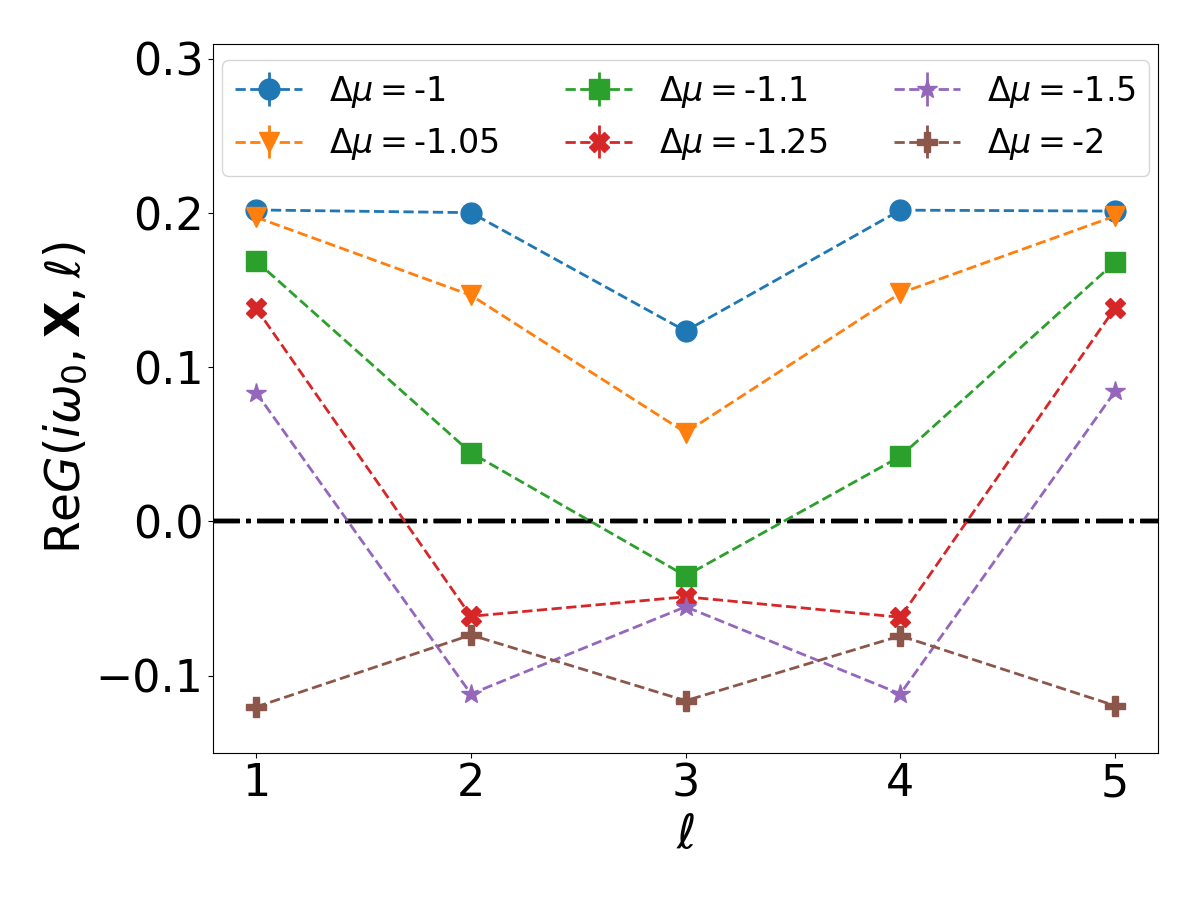} 
    \includegraphics[width=0.45\textwidth]{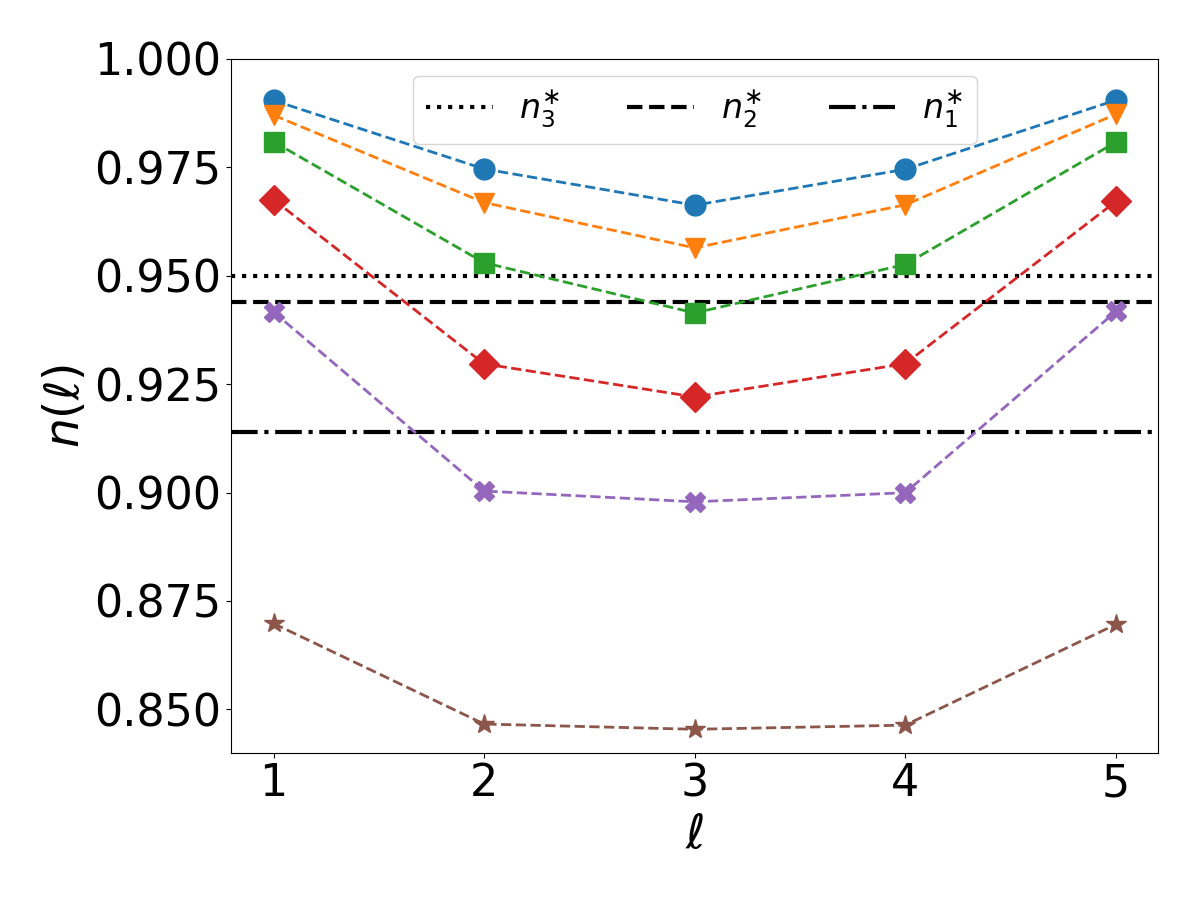}} 
    \caption{Left panel: Layer-dependent real part of the Green's function in patch $\mathbf{X}$ at the first Matsubara frequency for different $\Delta \mu=\mu-U/2$. The black dot-dashed line indicates the value at which the Fermi surface changes topology. We note that Fermi surfaces with different topology can coexist within the same slab. Right panel: Layer-dependent density profile upon changing $\Delta \mu$ as in the left panel. The horizontal lines represent the densities $n^*_{\ell}$ at which the Lifshitz transition occurs for the first 3 layers.}
    \label{fig:reg_5L and fig:n5_L}
\end{figure*}
\begin{figure}[hbt]
    \centering
    \includegraphics[width=0.45\textwidth]{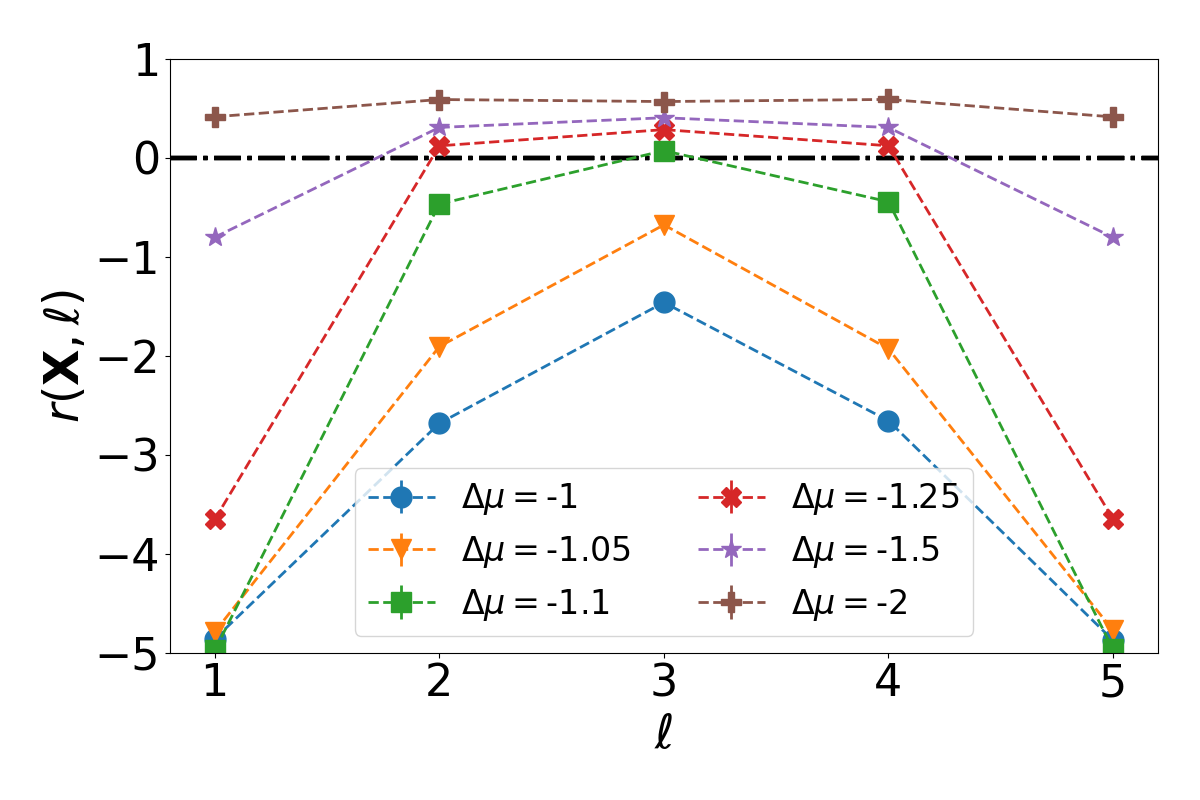}  
            \caption{Layer dependent renormalized dispersion \eqn{rk-layer}. The black line indicates the value of the Lifshitz transition. Although it is a fundamentally different quantity from $\Rea G(i \omega_0,\mathbf{X},\ell)$, $r(\mathbf{X},\ell)$ almost always gives the same prediction for the Fermi surface character.
            }
    \label{fig:rkl}
 \end{figure}

\noindent
As we already mentioned, weakly doping with holes the paramagnetic Mott insulating state of the 2D 
Hubbard model leads to a pseudogap phase characterized by a coexistence of a Luttinger surface and hole-like Fermi pockets~\cite{PhysRevB.74.125110,Georges-PRX2018,PhysRevLett.120.067002}. Upon further doping, the Fermi surface seems to undergo a Lifshitz transition and to turn into the expected electron-like one. We illustrate this scenario in Fig.~\ref{fig:hole_vs_electron}.  Concurrently, the pseudogap disappears and conventional Fermi liquid properties are recovered \cite{Georges-PRX2018,PhysRevLett.120.067002}.
\noindent
In a slab geometry, correlation effects are enhanced at the surface \cite{dead-layer,Antonio-dead-layer}, 
which envisages the possibility of a layer dependent Fermi surface topology.\\
We address this issue by studying a 5-layer doped Mott insulating slab for different values of $\Delta \mu=\mu- U/2 < 0$, which is the deviation of the chemical potential from the particle-hole symmetric point, $\Delta\mu=0$. Throughout this work, we take $U = 8$ and temperature $T = 1/40$, parameters for which we have verified that the Lifshitz transition coincides with the destruction of the pseudogap, see Appendix A for details. In the 2D Hubbard model, the order of the doping-driven transition from the pseudogap to the normal metal state appears to depend on the cluster size and method. Using cDMFT with a $2 \times 2$ plaquette, a first-order transition is observed \cite{Sordi-PRB}, whereas a crossover is found when using a 16-site cluster \cite{centre-focus}. Moreover, a first-order transition has not been reported in DCA calculations with clusters of various sizes \cite{PhysRevB.82.155101}.
In our slab system and for the set of parameters investigated in this work, we do not observe any clear first-order behavior when the pseudogap disappears. We further mention that the 
layers exhibiting a pseudogap and a hole-like Fermi surface have a non-Fermi liquid behavior of the 
self-energy, see Appendix B for details. \\ In the isolated, single-layer Hubbard model, the Lifshitz transition is conventionally characterized through the sign of the renormalized dispersion $r(\bk)=\ep(\bk)-\mu+\Rea\,\Sigma(i \omega_{0}, \bk)$, where $\omega_0 =\pi T$ is the first fermionic Matsubara frequency, computed at $\bk=\mathbf{X}$ \cite{FerreroDiagMC_PG,Georges-PRX2018,PhysRevLett.120.067002}. In a slab, since nearest neighbors layers are coupled to each other, the nature of the layer-resolved Fermi surface may be characterized through the real part of the layer-dependent Green's function at $\mathbf{X}$, $\Rea\, G(i \omega_0,\mathbf{X},\ell)$. When $\Rea\, G(i \omega_0,\mathbf{X},\ell)$ is positive(negative), the layer Fermi surface is hole(electron)-like. In Fig.~\ref{fig:reg_5L and fig:n5_L}, left panel, we plot this quantity for different values of $\Delta \mu$. We observe that $\Rea\, G(i \omega_0,\mathbf{X},\ell)$ strongly depends on the distance from the surface. In particular, the transition to an electron-like Fermi surface occurs at first in the central layer when $\Delta \mu\lesssim -1.1 $, while the outer layers, which feel an enhanced correlation 
strength, still display the hole-like Fermi surface of a weakly doped Mott insulator.
Decreasing $\Delta \mu$, we observe that the transition spreads to the superficial layers until it reaches the surfaces, $\ell=1$ and $N$, at $\Delta \mu\simeq -2 $.
In Fig.~\ref{fig:reg_5L and fig:n5_L}, right panel, we plot the layer dependent electron density for the same values of $\Delta\mu$ as in the left panel. We note that, at equal $\Delta \mu$, the outer layers are less doped that the inner ones, revealing once again the dead-layer effect.  
In the figure, we also show the density $n^*_{\ell}$ at which the Fermi surface reconstruction occurs in layer $\ell$. As expected, $n^*_{\ell}$ is significantly smaller at the surface layers, as if they 
were characterized by a larger $U$ \cite{PhysRevLett.120.067002,FerreroDiagMC_PG}.
For completeness, we also define a layer-dependent renormalized dispersion in the patch $\mathbf{X}$ through 
\be
r(\mathbf{X},\ell)=\ep(\mathbf{X})-\mu+\Rea\,\Sigma_{\ell}(i \omega_{0}, \mathbf{X})\,,
\label{rk-layer}
\ee
which we plot in Fig.~\ref{fig:rkl}. We expect that $r(\mathbf{X},\ell)>0$ corresponds to an electron-like Fermi surface, while 
$r(\mathbf{X},\ell)<0$ to a hole-like one. Indeed, the behavior of $r(\mathbf{X},\ell)$ in 
Fig.~\ref{fig:rkl} and that of $\Rea\,G(i\omega_0,\mathbf{X},\ell)$ in Fig.~\ref{fig:reg_5L and fig:n5_L}, left panel, 
are perfectly consistent with each other.
\subsection{Another marker of the Fermi surface reconstruction}
The hole-like Fermi surface at weak doping evidences a violation of Luttinger's theorem \cite{Luttinger}. The total number of electrons $N(T)$ at fixed $\mu$ and as function of temperature $T$ can be calculated through 
\beal 
N(T) &= -\sum_{\bk\sigma}\,T\sum_n\, \esp{i\omega_n 0^+}\;\fract{\partial \ln G(i\omega_n,\bk)}{\partial   i \omega_n}\\
&\qquad + \sum_{\bk\sigma}\,T\sum_n\, G(i\omega_n,\bk)\;\fract{\partial \Sigma(i\omega_n,\bk)}{\partial i\omega_n}\\
&= N_L(T) + I_L(T)\,.
\label{IL}
\eal
At $T=0$, the number of electrons is an integer. Also $N_L(0)$ becomes an integer and corresponds 
to twice, because of spin, the number of the $\bk$-points in the regions enclosed by the surfaces at which the Green's function at zero frequency $G(0,\bk)$ changes sign. We emphasize that such sign change occurs at the Fermi surface, where $G(0,\bk)$ has a pole, as well as at the Luttinger surface \cite{Igor-PRB2003}, where $G(0,\bk)$ vanishes. In other words, both surfaces should contribute to the electron counting if Luttinger's theorem, which states that $I_L(0)=0$, were literally taken. However, it has been demonstrated \cite{Heath_2020} that Luttinger's theorem is violated in presence of a 
Luttinger surface. In such case, therefore, $I_L(0)$ must be not only finite but also quantized in  integer values. We recall that Luttinger's demonstration of his theorem invokes the invariance of 
the Luttinger-Ward functional with respect to a global shift of the Green's function frequencies, 
which implies that, e.g., 
\be
I^\Delta_L(T)=\sum_\bk\,T\sum_n\, G(i\omega_n,\bk)\;\Delta \Sigma(i\omega_n,\bk)=0\,,
\label{DIL}
\ee
where 
\bealn
\Delta \Sigma(i\omega_n,\bk) =\fract{\Sigma(i\omega_n+i\Omega,\bk)-\Sigma(i\omega_n-i\Omega,\bk)}{2i\Omega}  \;,
\eal
with $\Omega=2\pi T$, is the discrete derivative. It is tempting to argue that, in the limit $T\to 0$,  
$\Delta \Sigma(i\omega_n,\bk) \to \partial_{i\omega}\,\Sigma(i\omega,\bk)$, thus that \eqn{DIL} 
implies $I_L(0)=I^\Delta_L(0)=0$. However, this conclusion is generally not true. The reason is 
that the function that is summed in $I^\Delta_L(T)$ of \eqn{DIL} has different   
analytic properties at any finite $T$ than the function in $I_L(T)$ of \eqn{IL}, recalling that the imaginary parts of the Green's function and self-energy may have a discontinuity crossing zero Matsubara frequency. In the supplementary material of Ref.~\cite{Michele2} it was demonstrated that at leading order of an expansion in powers of $T$ 
\beal
&I_L(T) - I^\Delta_L(T) =  I_L(T) \\
&\; \simeq -\fract{1}{\pi}\,\sum_\bk\, 
\Rea\,G(i\pi T,\bk)\,\Ima\,\Sigma(i\pi T,\bk) + O\big(T^2\big)\,,
\label{IL expansion}
\eal
where the first term on the right hand side is linear in $T$ both in absence and in presence of 
a Luttinger surface.\\
In Ref.~\cite{Jan-PRB2022} it was further shown that $I_L(T)$ is a boundary term, actually the 
value divided by $\pi$ of the imaginary part of a function    
evaluated at the first Matsubara frequency $\omega_0=\pi T$. Since $I_L(T\to 0)$ is an integer, 
it was also conjectured \cite{Jan-PRB2022} that such imaginary part is the phase accumulated by a complex function $\Phi(i\omega_n)=\Phi(-i\omega_n)^*$ from 0 to $\infty$, and thus that 
\be
I_L(T) = \fract{1}{\pi}\,\tan^{-1}\fract{\Ima\,\Phi(i\omega_n)}{\Rea\,\Phi(i\omega_n)}
_{\big|^{\infty}_{\omega_0}}
\;,
\label{IL Jan}
\ee
with $\Ima\,\Phi(i\omega_n) \sim \omega_n$ at small $\omega_n$, consistently with \eqn{IL expansion}.
\begin{figure}[hbt] 
\centerline{
\includegraphics[width=0.45\textwidth]{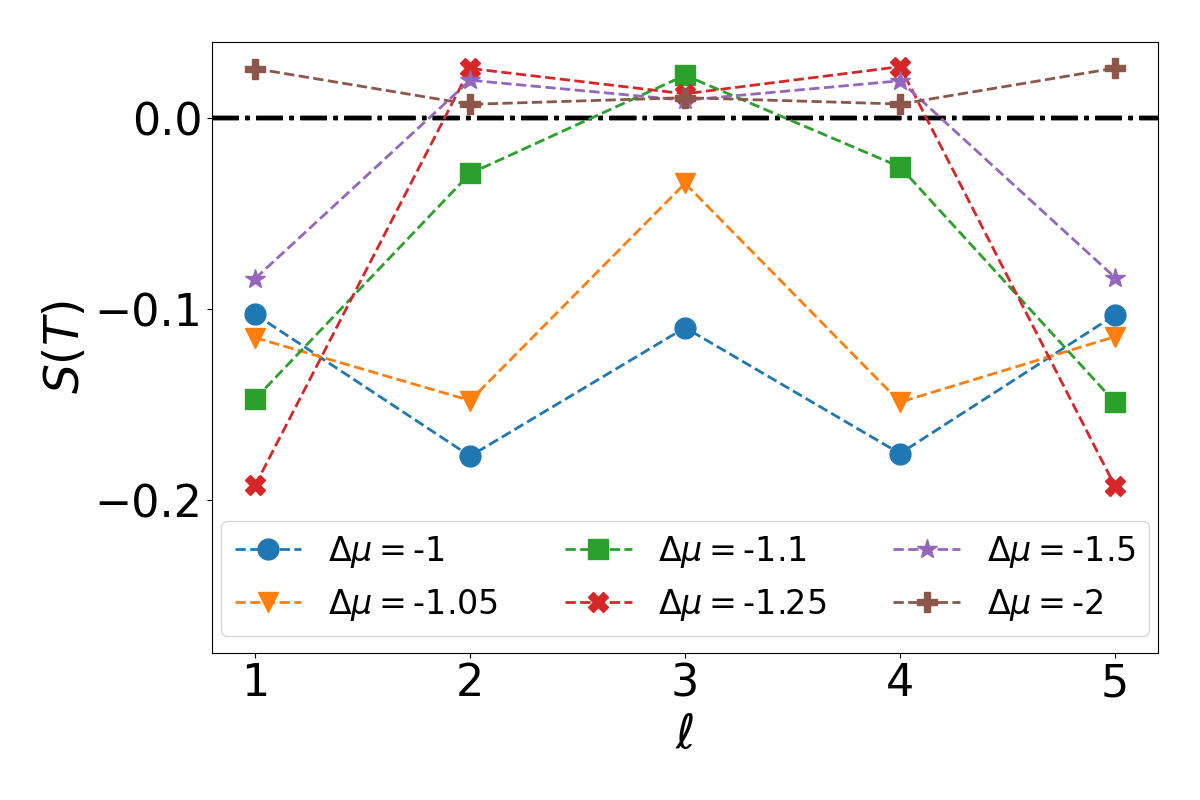} 
}
\caption{Layer dependence of $S(T)$ in \eqn{ST} for different values of $\Delta\mu$.}
\label{S(T)}
\end{figure}
The proposed expression \eqn{IL Jan} shows that, despite the expansion in $T$ suggests that 
$I_L(T\to 0)\to 0$, in reality the series converges to a finite integer 
value if $\Rea\,\Phi(i\omega_0\to 0)<0$. 
If we assume that (8) is correct, thus that \eqn{IL expansion} is actually the leading term of the expansion in powers of $T$ of $\text{arg}\big(\Phi(i\pi T)\big)$, we may diagnose Luttinger's theorem breakdown simply by the sign change of 
\be
S(T) = \sum_\bk\, 
\Rea\,G(i\pi T,\bk)\,\Ima\,\Sigma(i\pi T,\bk)\,. 
\label{ST}
\ee
In Fig.~\ref{S(T)} we plot the layer dependence of $S(T)$ for the same values of $\Delta\mu$ of  
Fig.~\ref{fig:reg_5L and fig:n5_L}. We observe that the sign change of $S(T)$ reproduces well 
the Lifshitz transition that we inferred from the figures \ref{fig:reg_5L and fig:n5_L} and   
\ref{fig:rkl}. This result not only gives support to the prediction \eqn{IL Jan} made in \cite{Jan-PRB2022}, but also demonstrate that $S(T)$ in \eqn{ST}, a quantity that can be calculated by DCA 
without any periodization, can be employed as an effective marker of the Luttinger's theorem breakdown 
due to the emergence of a Luttinger surface.

\section{Friedel oscillations} \label{Friedel-oscillations}
The Friedel oscillations (FOs) of the electron density or of the density of states occur in metals due to the presence of defects \cite{Friedel1,Friedel2}, regarded as sources of translational 
symmetry breaking with dimension $d_*<d$, $d$ being the system dimension. For instance, 
an impurity corresponds to a point-like defect with $d_*=0$, a dislocation in a crystal to a line defect with $d_*=1$, and an interface to a plane defect with $d_*=2$.\\
For an isotropic Fermi surface, the FOs decay asymptotically as a power law in the distance $r$ from the defect \cite{Kittel}, 
\beal
\text{FOs}(r) & \sim \fract{\cos(2 k_F r+\delta)}{r^{d-d_*}}\;,& k_F\,r&\gg 1\,,
\label{eqn:generic_FO}
\eal
where $k_F$ is the Fermi momentum and $\delta$ a phase shift. In our slab geometry, the FOs are caused by the surfaces, thus $d_*=2$ and $d=3$. Since the periodicity depends on $k_F$, the FOs are very sensitive to the shape and topology of the Fermi surface \cite{FO-weyl}.  In this Section, we investigate the signatures of the doping-driven Fermi surface reconstruction on the FOs caused by the open geometry. 

\subsection{Friedel oscillations in a non-interacting slab}  \label{FO-non-int}
\begin{figure}[h!]
    \centerline{
    \includegraphics[width=0.5\textwidth]{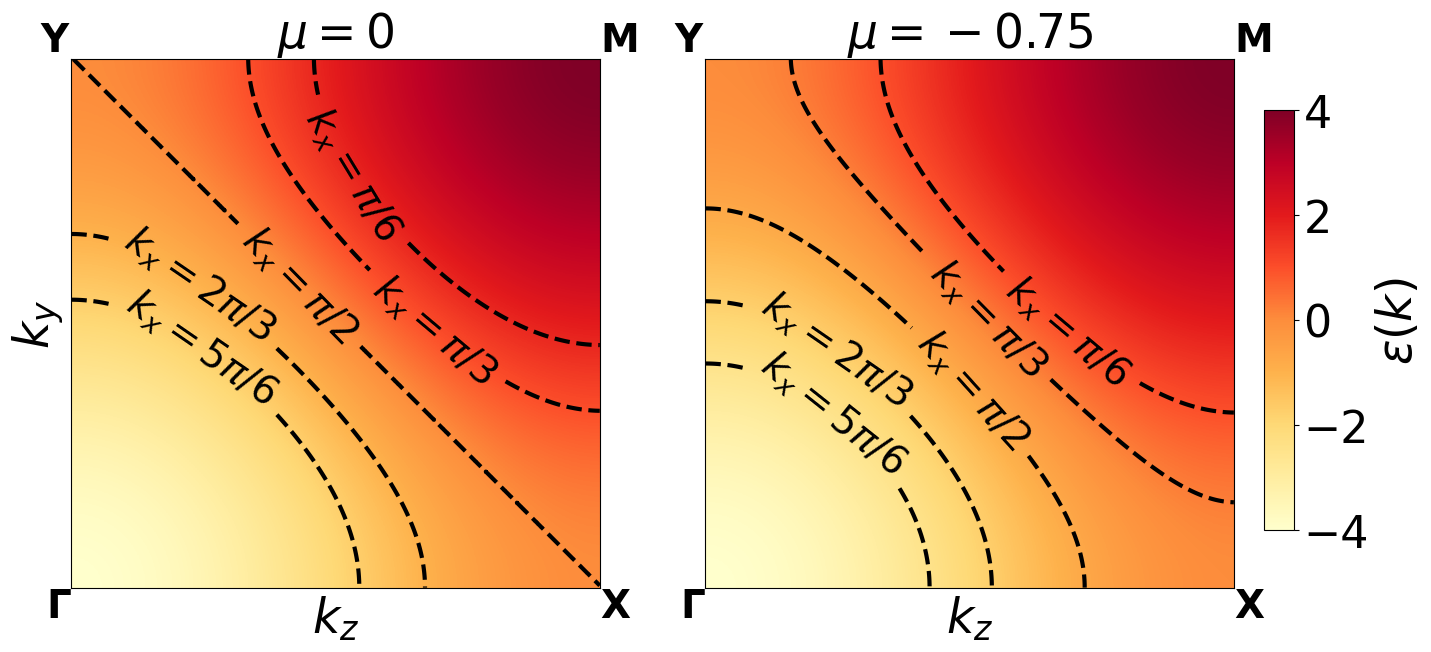}}
            \caption{(a) Color map of the in plane dispersion $\ep(\bk)$ in the upper right quarter of the Brillouin zone.  The dashed lines represent the Fermi sheets for each value of $k_{x}$, see \eqn{sheets}, in a 5-layer non-interacting half-filled slab. (b) Same as (a) but for $\mu=-0.75$, where the slab is hole-doped. }
    \label{fig:curve_di_livello_FS}
\end{figure}

\begin{figure*}[ht]
        \centerline{
       \includegraphics[width=0.3\textwidth]{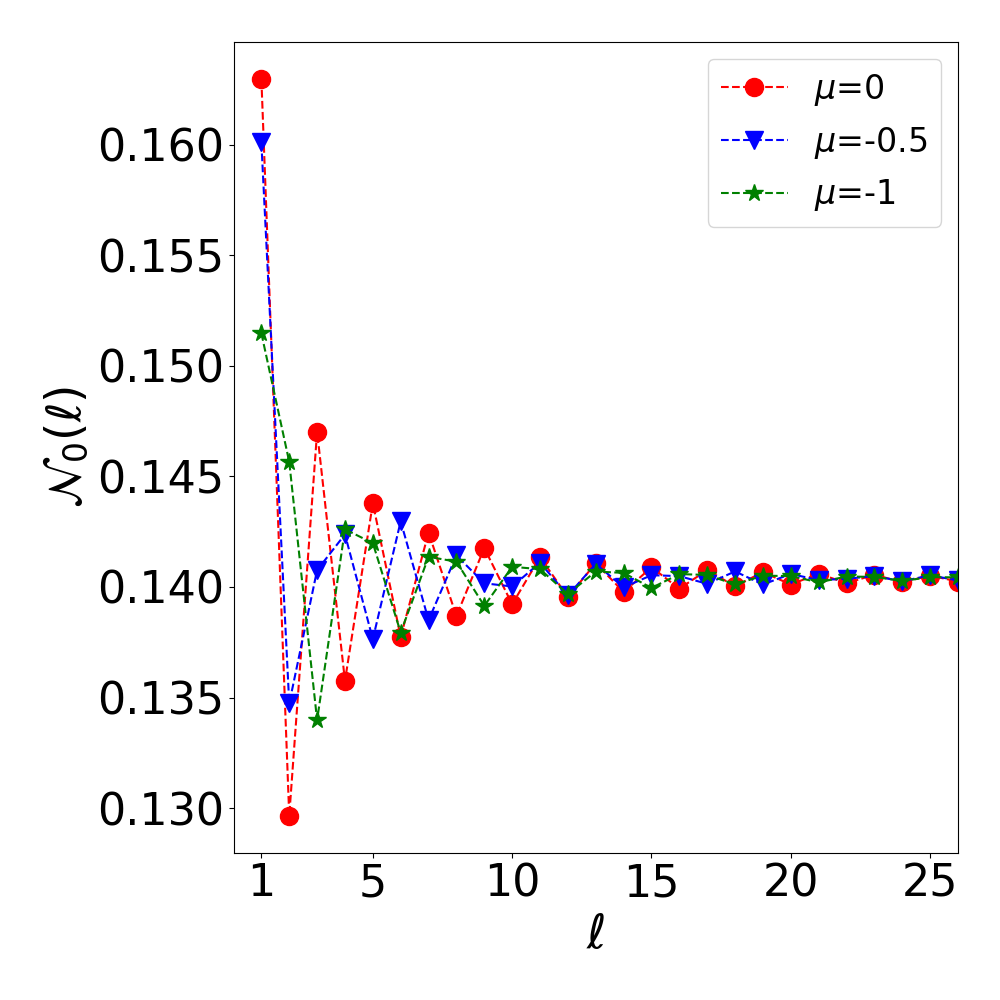}
            \includegraphics[width=0.3\textwidth]{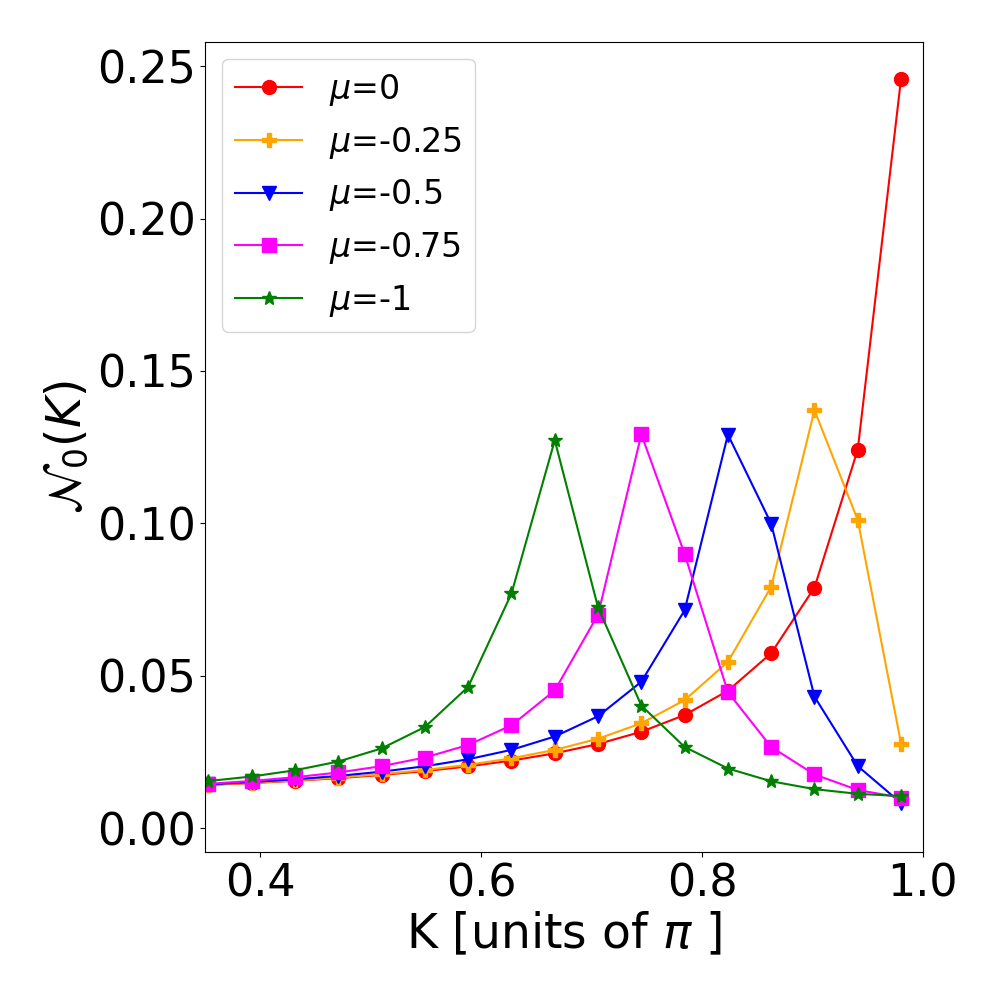}
            \includegraphics[width=0.3\textwidth]{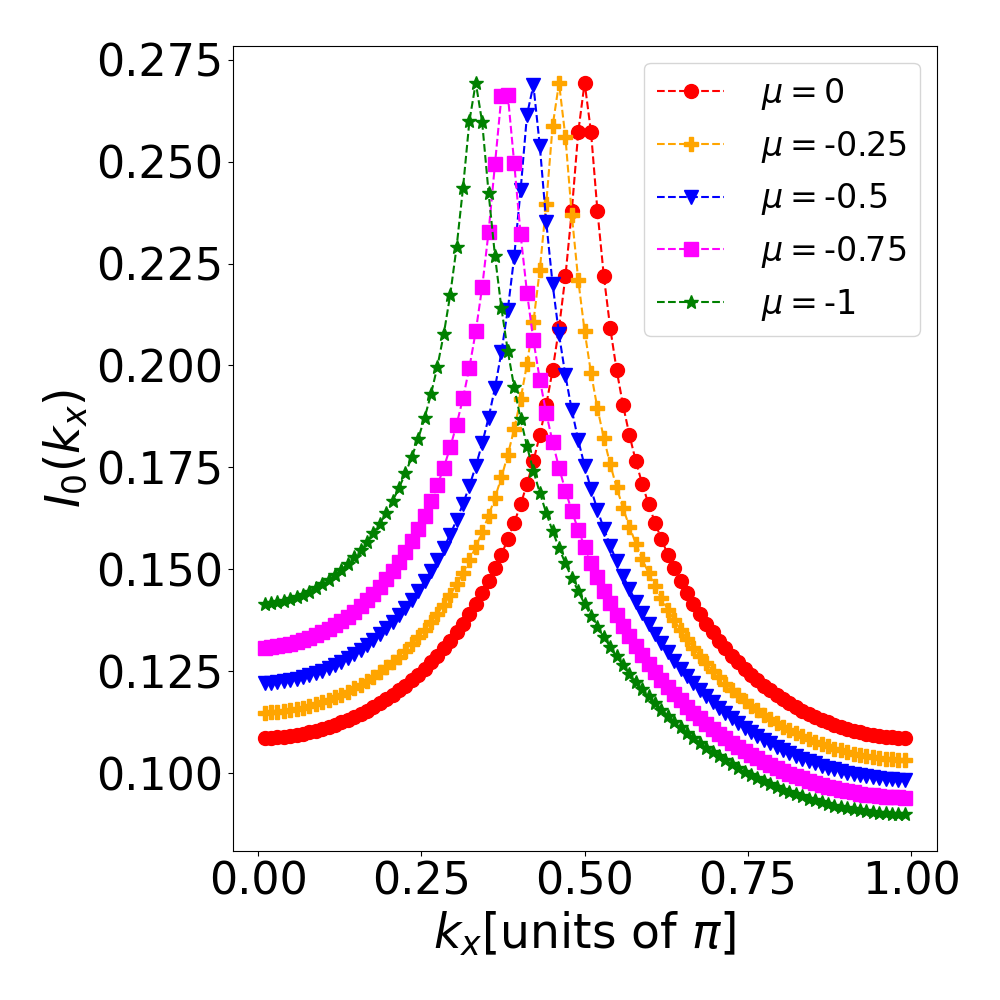}}
        \caption{Left panel: Profile of $\mathcal{N}_0(\ell)$ in a non-interacting slab with $N=101$. We note the $\mu$-dependence of the FOs period. Middle panel: Discrete Fourier transform $\mathcal{N}_0(K)$ of $\mathcal{N}_0(\ell)$, showing a well defined peak for every value of $\mu$. In the case of $\mu=0$ the peak is truncated by the aliasing. Right panel: $I_0(k_x)$ of \eqn{eqn:I_def} for different values of $\mu$. The position of the maximum of $I(k_x)$ is denoted as $K_M$. }
        \label{fig:friedel_non_int}
    \end{figure*}

In order to have a reference case, we start by considering our slab in absence of interaction. 
In this case, the electron wavefunction is the product of an in-plane Bloch wave $\psi_{\bk}(\bR)$ and out-of-plane one, 
\beal
\phi_{k_x}(\ell)&= \sqrt{\frac{2}{N+1}\;}\; \sin k_x \ell\,,& k_x&= \fract{\pi n_x }{N+1}\;,
\eal
where both $n_x$ and $\ell$ run from 1 to $N$. 
The Green's function in the diagonal basis simply reads
\be
G_0(i\omega_n,\bk,k_x)=\fract{1}{\;i \omega_n+\mu-\ep(\bk)+2t \cos k_x\;}\;, 
\label{eqn:G_kx}
\ee
thus the layer-dependent local Green's function 
\be
G_0(i\omega_n,\ell) =\sum_{k_x}\,\phi_{k_x}^{2}(\ell)\,\fract{1}{V}\sum_\bk\,G(i\omega_n,\bk,k_x)\,,
\label{eqn:G_ell}
\ee
with $V$ the number of sites within each layer.
We here concentrate on the FOs in the local spectral function at the chemical potential, $\mathcal{N}(\ell)$, which can be approximated by the imaginary part of the local Green's function at the first Matsubara frequency, $\mathcal{N}(\ell)\simeq -\Ima\, G( i \omega_0,\ell)/\pi$  \cite{Fate_pg,PhysRevLett.131.166501,deng2019signatures}.
The accuracy of this approximation improves at low temperatures, as we have verified 
comparing the results at the first Matsubara frequency with those extrapolated at zero frequency. We choose $\mathcal{N}(\ell)$ because the FOs in the density are severely suppressed by correlations, and thus harder to disentangle from the QMC noise. In the non-interacting slab, 
\beal
\mathcal{N}_0(\ell) &= -\fract{1}{\pi} \,\Ima\, G_0( i \omega_0,\ell) = \sum_{k_x}\, \phi_{k_x}^{2}(\ell)\, I_0(k_x)\,,\\
I_0(k_x) &=-\fract{1}{\pi V} \sum_\bk\, \Ima\,G_0(i\omega_0,\bk,k_x)\,.
\label{eqn:I_def}  
\eal
Equation \eqn{eqn:I_def} allows quantifying through $I_0(k_x)$ the contributions to 
$\mathcal{N}_0(\ell)$ of the different $k_x$-dependent Fermi surface sheets, $\bk_F(k_x)$, defined through 
\be
\ep\big(\bk_F(k_x)\big) = \mu + 2t\,\cos k_x\,.
\label{sheets}
\ee
In Fig.~\ref{fig:curve_di_livello_FS} we plot these sheets in a slab with $N=5$, for both the half-filled and hole-doped cases.  
In Fig.~\ref{fig:friedel_non_int}, left panel, we plot $\mathcal{N}_0(\ell)$ for a slab with $N=101$ at different values of chemical potential. We observe that the period of the FOs depends on the value of $\mu$, which is to be expected since lowering the chemical potential decreases the size of the Fermi surface. We quantitatively investigate the period by studying the discrete Fourier transform 
$\mathcal{N}_0(K)$ of $\mathcal{N}_0(\ell)$. To isolate the oscillations, we subtract the bulk value and assume a $1/\ell$ decay, consistently with \eqn{eqn:generic_FO}. The Fourier spectra are plotted in the middle panel of Fig.~\ref{fig:friedel_non_int} and each show a sharp peak that moves with $\mu$. In the right panel of Fig.~\ref{fig:friedel_non_int} we instead plot $I_0(k_x)$ of \eqn{eqn:I_def}. This function shows a sharp peak, whose position we denote as $K_{M}$. We find that $2K_M$ compares well with the position of the peak in $\mathcal{N}_0(K)$ for each value of $\mu$. This suggests that the sum over $k_x$ in \eqn{eqn:I_def} can be faithfully computed by the maximum of 
$I_0(k_x)$. Numerically, $K_{M}$ is found to be related to $\mu$ by 
\beal
\cos K_{M} \simeq -\fract{\mu}{2t}\;.
\eal 
Therefore, and unsurprisingly, the maximum of $I_0(k_x)$ is at the value of $k_x$ that 
maximizes the corresponding Fermi surface sheet \eqn{sheets}, which corresponds, in our case, to the 
half-filled 2D Fermi surface.  
\subsection{Friedel oscillations in the interacting slab} \label{FO-int}
\begin{figure}[ht]
    \centering
    \includegraphics[width=0.45\textwidth]{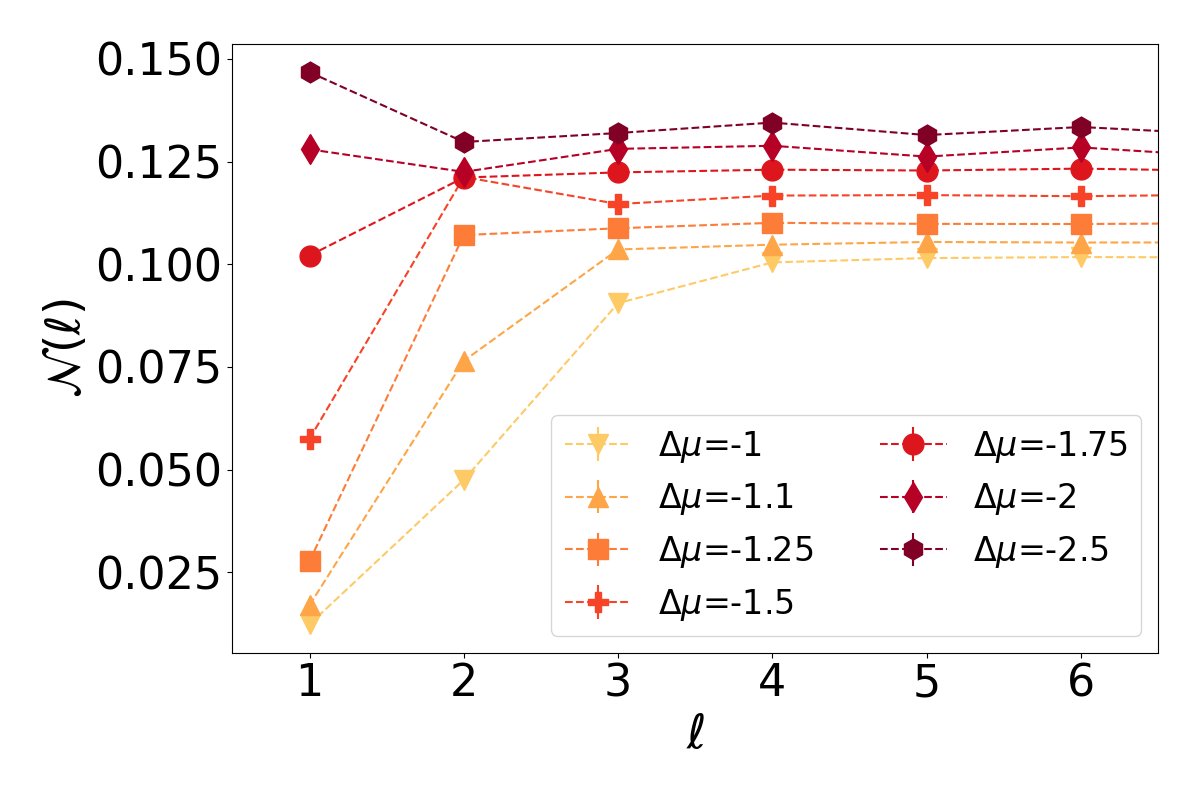} 
            \caption{Spectral density $\mathcal{N}(\ell)$ for different values of $\Delta \mu$ in an $N=11$ interacting slab. As doping grows, $\mathcal{N}(\ell)$ has a crossover from a surface dead-layer like profile to a profile showing Friedel oscillations.  }
    \label{fig:Friedel-fully-scf}
 \end{figure}

In the interacting slab, $k_x$ is no longer a good quantum number and the analysis of the FOs 
is not as simple as in the absence of interaction. 
In Fig.~\ref{fig:Friedel-fully-scf} we plot $\mathcal{N}(\ell)$ for a $N=11$, which we obtain by averaging $ -\fract{1}{\pi}\,\Ima G( i \omega_0, \mathbf{P},\ell)$ over the four patches. We observe the absence of the FOs until $\Delta \mu$ becomes smaller that $\simeq -1.1$, though 
the slab is doped away from half-filling. 
In this regime of missing Friedel oscillations, $\mathcal{N}(\ell)$ increases monotonously from the edges to the center of the slab, a further manifestation of the surface dead-layer effect. 
However, as soon as the FOs appear, which happens for $\Delta\mu\lesssim -1.1$, $\mathcal{N}(\ell)$ stops being monotonous. In other words, 
$\mathcal{N}(\ell)$ shows a crossover from a regime dominated by the dead-layer effect to one where 
more conventional FOs emerge. 
Another fingerprint of this crossover is the behavior of the number of layers $\ell_\text{bulk}$ necessary to reach the bulk value of $N(\ell)$.
In the regime where the dead-layer dominates, increasing the number of holes decreases 
$\ell_\text{bulk}$, since the system is moving further away from the Mott insulator and the correlation length decreases \cite{dead-layer,Antonio-dead-layer}.
In contrast, in the conventional FOs regime, the same quantity increases with hole-doping, as FOs are less suppressed by correlation.

\subsection{Friedel oscillations in a slab with homogeneous self-energy} \label{FO-HSEA}
\begin{figure*}[t!]
\centerline{
        \includegraphics[width=0.38\textwidth]{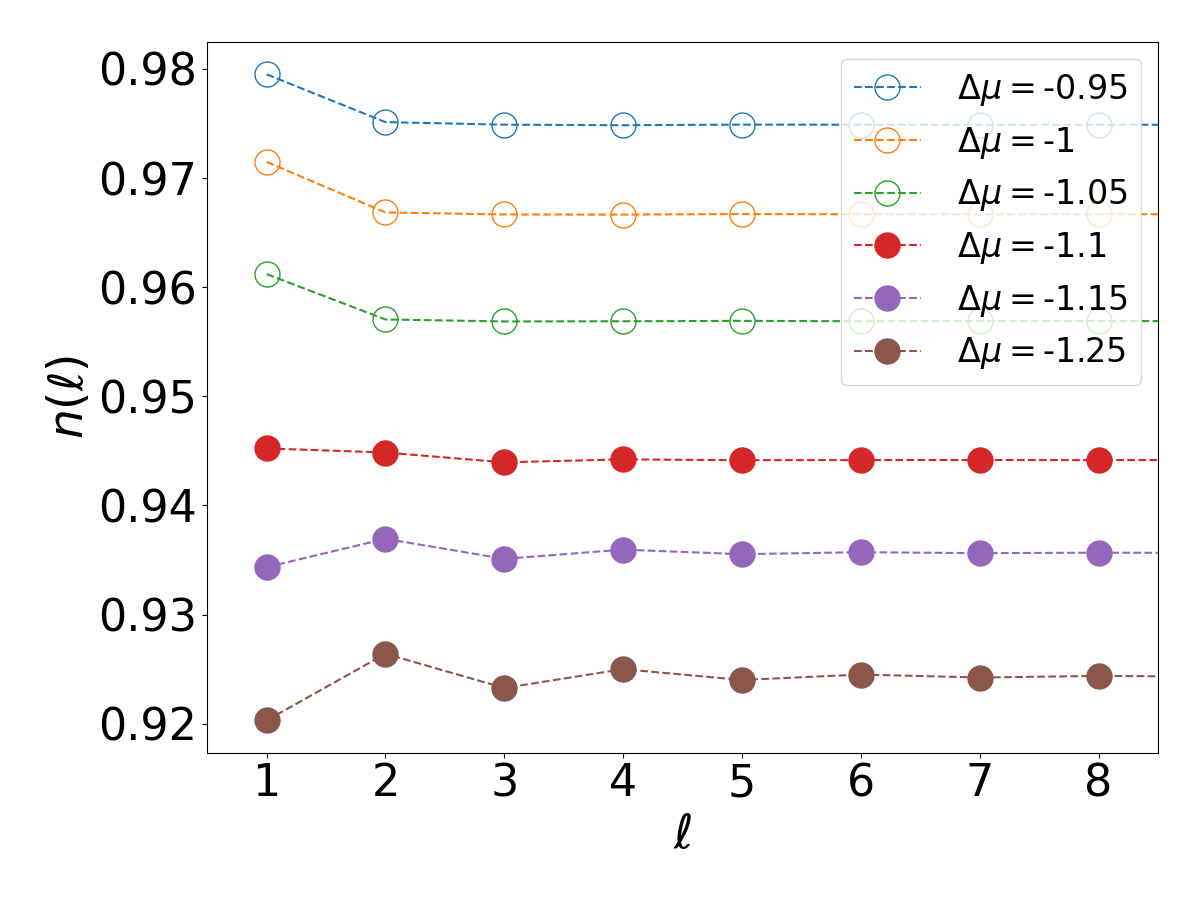}
        \includegraphics[width=0.38\textwidth]{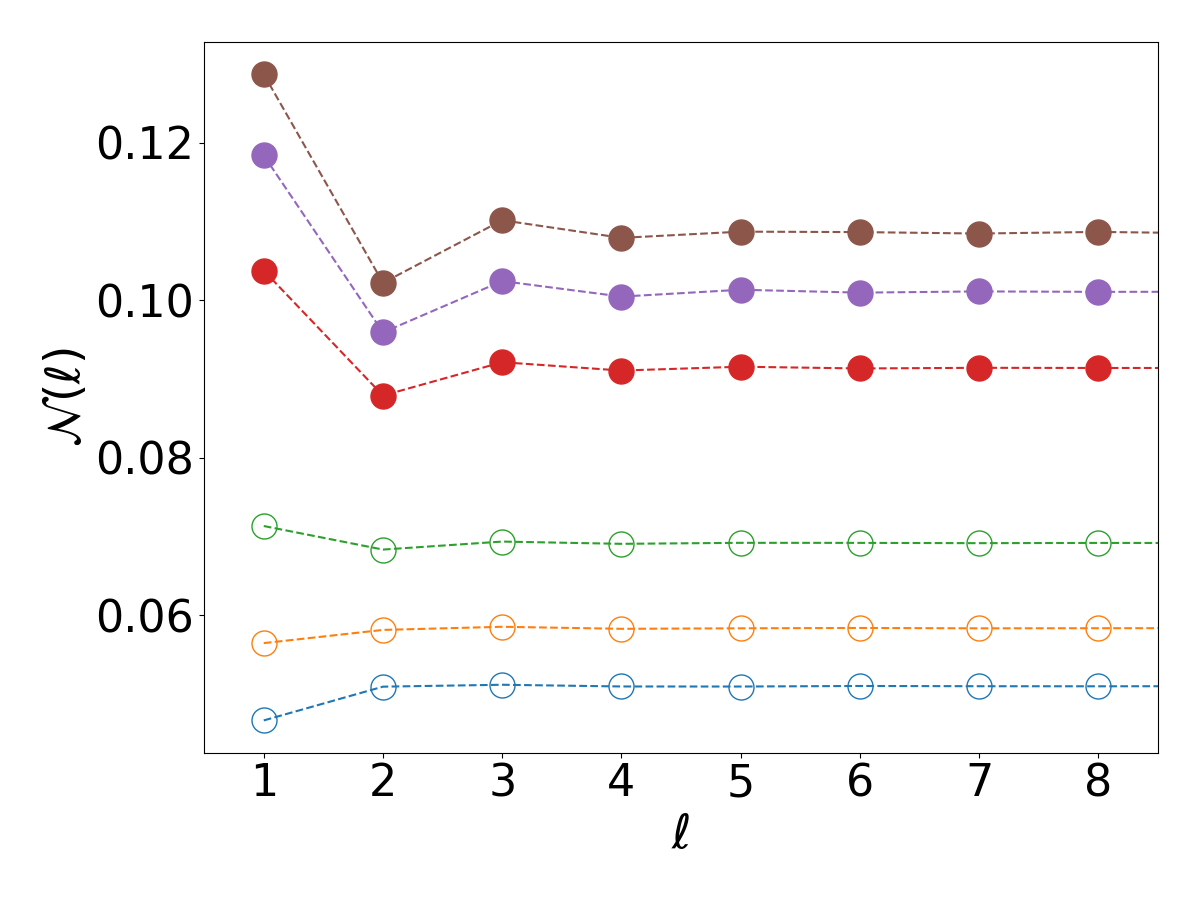}
        }
    \caption{Density, $n(\ell)$ in left panel, and spectral weight at the chemical potential, $\mathcal{N}(\ell)$ in right panel, profiles of a slab with $N=101$ computed with a layer-independent self-energy. The empty(filled) dots represent slabs with hole(electron)-like Fermi surface in the layers. In both quantities the oscillations are strongly suppressed when the Fermi surface has hole-like character. Instead, the qualitative profile when the Fermi surface is electron-like resembles the non-interacting one. }
    \label{fig:density-HSEA and fig:IMG-HSEA}
\end{figure*}

\begin{figure}[t!]
    \centerline{
            \includegraphics[width=0.23\textwidth]{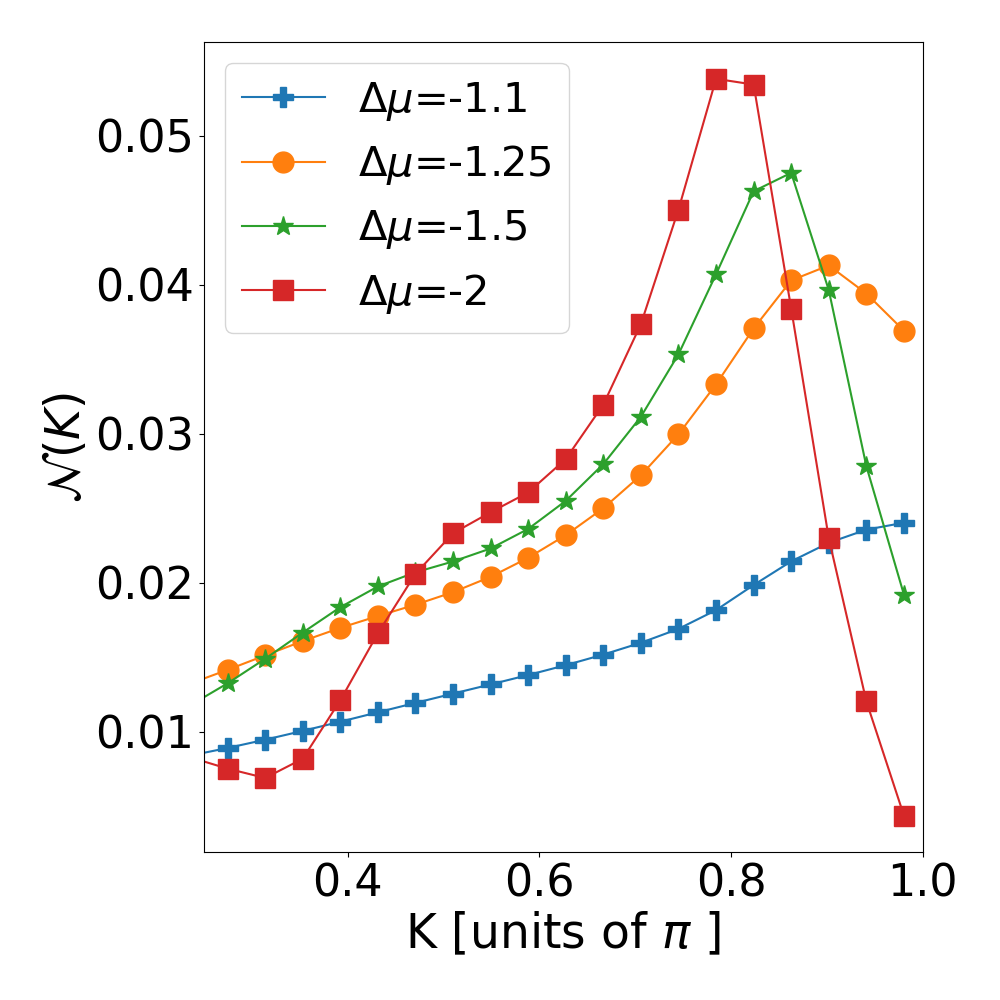}
            \includegraphics[width=0.23\textwidth]{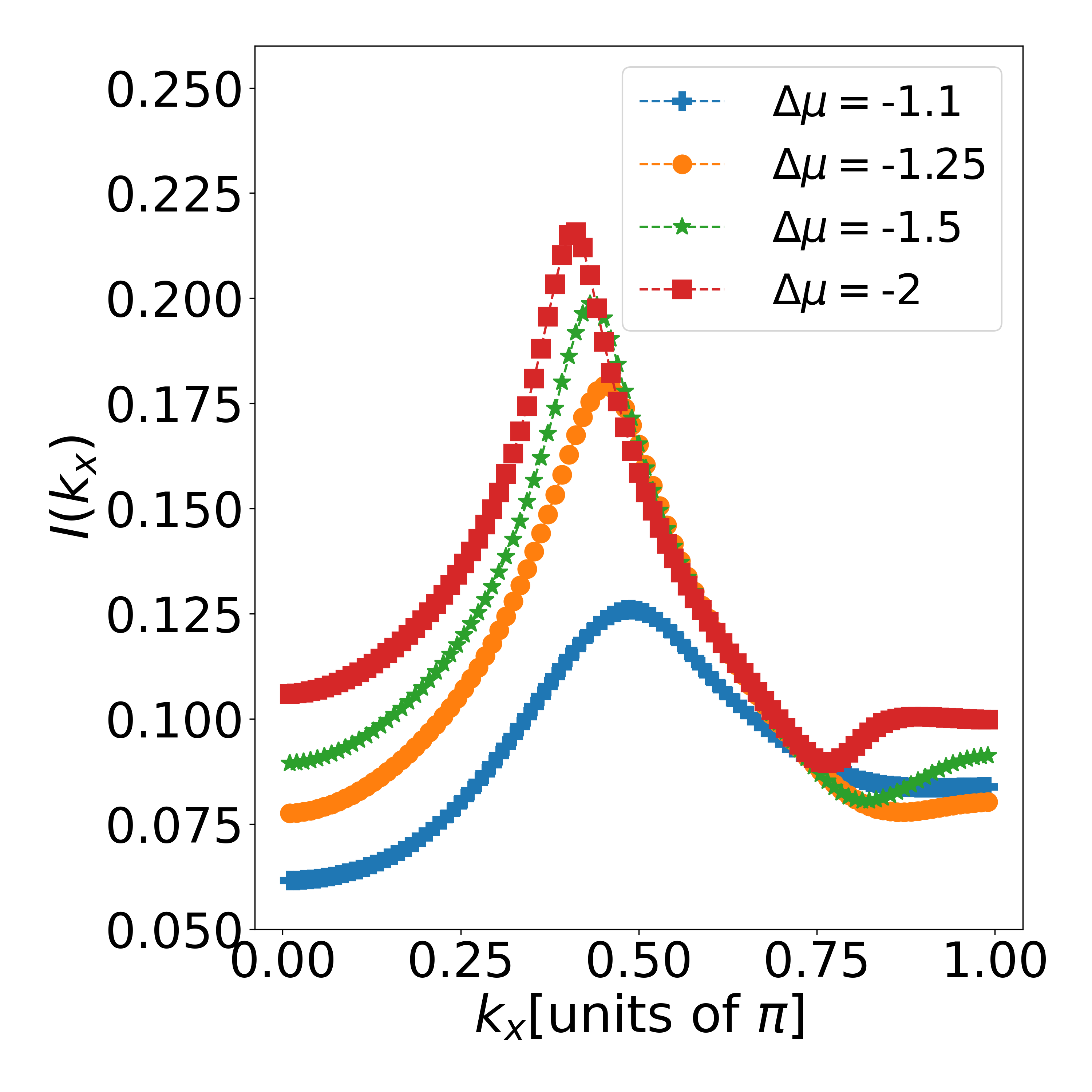}
            }
        \caption{Left panel: Discrete Fourier transform of $\mathcal{N}(\ell)$ for  a slab with $N=101$ where all the layers share the same electron-like Fermi surface. When doping is increased the position of the peak decreases and the peak becomes sharper and more intense. Right panel: $I(k_x)$ for the same values as in left panel. }
        \label{fig:FFT-e-integrale-electron-like}
\end{figure}

The Friedel oscillations are evidently sensitive to the size and shape of the Fermi surface. Therefore, they should bear signatures of the Lifshitz transition we earlier discussed, specifically of the fact that at low doping the Fermi surface is composed of hole-like Fermi pockets and turns electron-like  
above a threshold doping. For that, we could perform the Fourier transform of $\mathcal{N}(\ell)$, in analogy to the non-interacting case. However, it is numerically not affordable to study a slab with a number of layer larger than 11, which is not enough to perform a meaningful Fourier analysis. Moreover, the coexistence of Fermi surfaces of different topologies makes the process of identifying the fingerprints of a specific Fermi surface character significantly harder. To overcome these issues, we study a simplified system where the self-energy is assumed not to depend on the layer index. A similar approach, known as homogeneous self-energy approximation (HSEA) \cite{Jan-HSEA} has been used to study the FOs caused by an impurity in a 2D Hubbard model for large system size. 
Since the dead-layer effects decay exponentially in the distance from the surface \cite{dead-layer}, this approximation is expected to reproduce the leading asymptotic behavior of the FOs deep in the bulk.
In our specific example, we implement HSEA by considering the DCA self-energy of the central layer ($\ell=3$) of the 5-layer slab, computed through the fully self-consistent algorithm described in Section \ref{LT}, and use it as a homogeneous self-energy for all layers of a slab with $N\gg 5$ at the same chemical potential as the original $N=5$ slab. In this way, all layers share the same Fermi surface topology. This method allows us to treat slabs with a large number of layers and thus study with a better resolution the Fourier transform of $\mathcal{N}(\ell)$. Moreover, since the self-energy is the same for the all layers, $k_x$ is again a good quantum number, which allows straightly generalizing \eqn{eqn:I_def} replacing $I_0(k_x)$ with 
\beal
I(k_x) &=  -\fract{1}{4\pi}\,\Ima\sum_{\mathbf{P}}\, \fract{1}{N_{\mathbf{P}}}\, \sum_{\bk \in \mathbf{P}} \\
&\;
\fract{1}{\;i \omega_0+\mu-\ep(\bk)+2t \cos(k_x)-\Sigma(i \omega_0,\mathbf{P})\;}\;. 
        \label{eqn:I_con_sigma}
        \eal
In the left panel of Fig.~\ref{fig:density-HSEA and fig:IMG-HSEA} we plot the density profile of a  slab with $N=101$ obtained within the HSEA for different values of $\Delta \mu$ across the Lifshitz transition. We observe that the oscillations are almost absent at low doping before the transition, where only the first layer is significantly affected, and reappear at large doping after the transition. In the right panel of Fig.~\ref{fig:density-HSEA and fig:IMG-HSEA} we instead plot $\mathcal{N}(\ell)$. It shows a similar behavior: when the Fermi surface is hole-like the oscillations are suppressed and $\mathcal{N}(\ell)$ is usually smaller at the edges. Conversely, when the Fermi surface becomes electron-like, the FOs have larger amplitude and $\mathcal{N}(\ell)$ at the surface is larger than in the bulk, similarly to what happens in the non-interacting case. This qualitative difference is not observed if one computes $\mathcal{N}(\ell)$ using a homogeneous single-site DMFT self-energy, in which case $\mathcal{N}(\ell)$ is always larger at the surface. This suggests that the observed difference is in fact related to the change in Fermi surface topology, which single-site DMFT cannot capture.
\noindent
To better address this question, we study the Fourier transform $\mathcal{N}(K)$ of $\mathcal{N}(\ell)$. In Fig.~\ref{fig:FFT-e-integrale-electron-like}, left panel, we show $\mathcal{N}(K)$ for $\Delta \mu$ such that the Fermi surface is electron-like. The peaks in the Fourier spectrum are much broader than in the non-interacting case, although they become sharper and sharper upon increasing doping. As in the non-interacting case, the position of the maximum in the spectrum decreases with increasing doping.\\
In Fig.~\ref{fig:FFT-e-integrale-electron-like}, right panel, we plot $I(k_x)$ in \eqn{eqn:I_con_sigma} 
for the values of $\mu$ that yield an electron-like Fermi surface. As in the non-interacting case, 
if $K_M$ is the position of the peak in $I(k_x)$, then $2K_{M}$ agrees with the peak position in $\mathcal{N}(K)$. Upon defining the renormalized chemical potential in patch $\mathbf{X}$ as $\mu_{\bx}=\mu-\Rea \Sigma(i \omega_0,\mathbf{X})$, $K_M$ is well approximated by 
\beal
\cos K_{M} =-\fract{\,\mu_{\bx}\,}{2t}\;.
\label{eqn:mux}
\eal
In other words, the conclusion in Sec.~\ref{FO-non-int} is still valid: $K_M$ is the value of $k_x$ that maximizes the Fermi surface sheet given by \eqn{sheets}, with the only difference that the chemical potential is the renormalized one. Another important difference with the non-interacting case is that $I(k_x)$ is significantly less peaked that its non-interacting counterpart. This is a consequence of the nonzero imaginary part of the self-energy that broadens the peak. This effect, which diminishes with doping, is reflected also in the Fourier spectrum of $\mathcal{N}(\ell)$. The relation \eqn{eqn:mux} also implies that the FOs period is predominantly dictated by the patch $\mathbf{X}$, which is the patch that contains the largest portion of the electron-like Fermi surface.\\

\subsection{Evidence of Fermi pockets from the Friedel oscillations at low doping}
\begin{figure}[t!]
    \centerline{
            \includegraphics[width=0.23\textwidth]{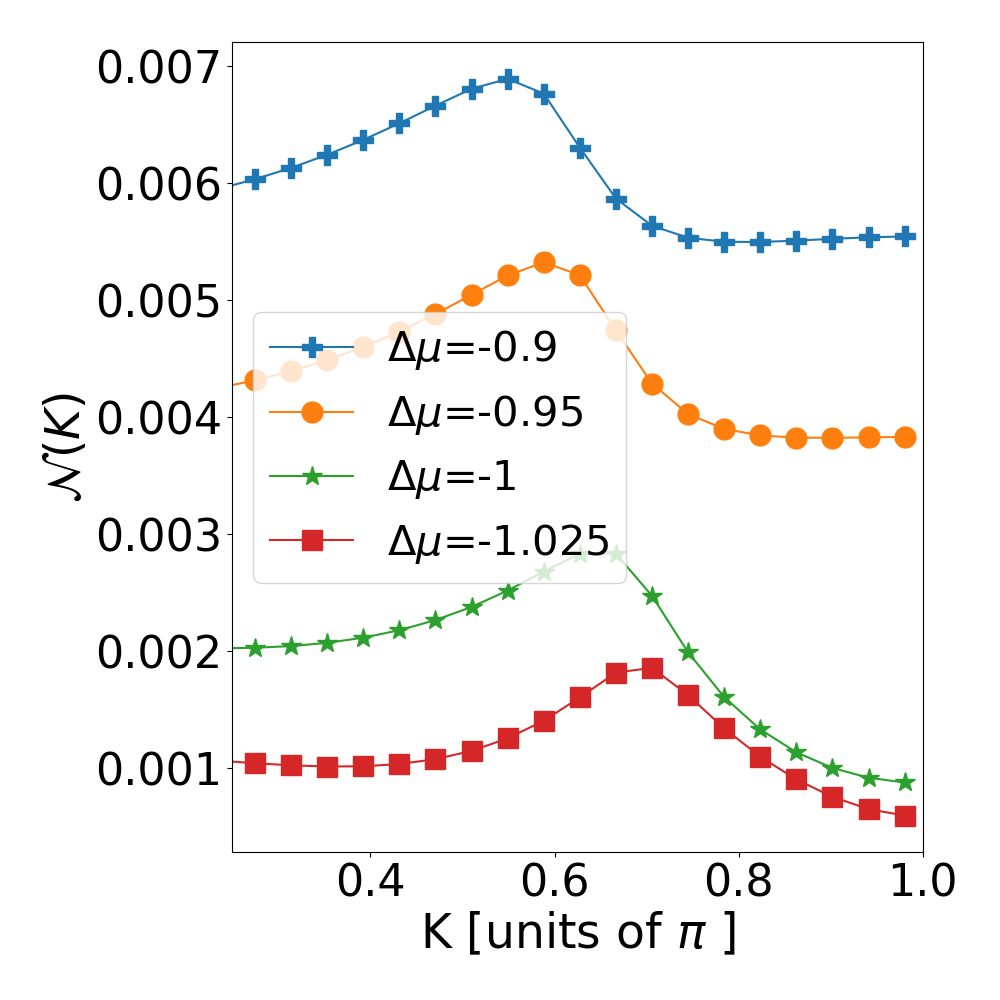}
            \includegraphics[width=0.23\textwidth]{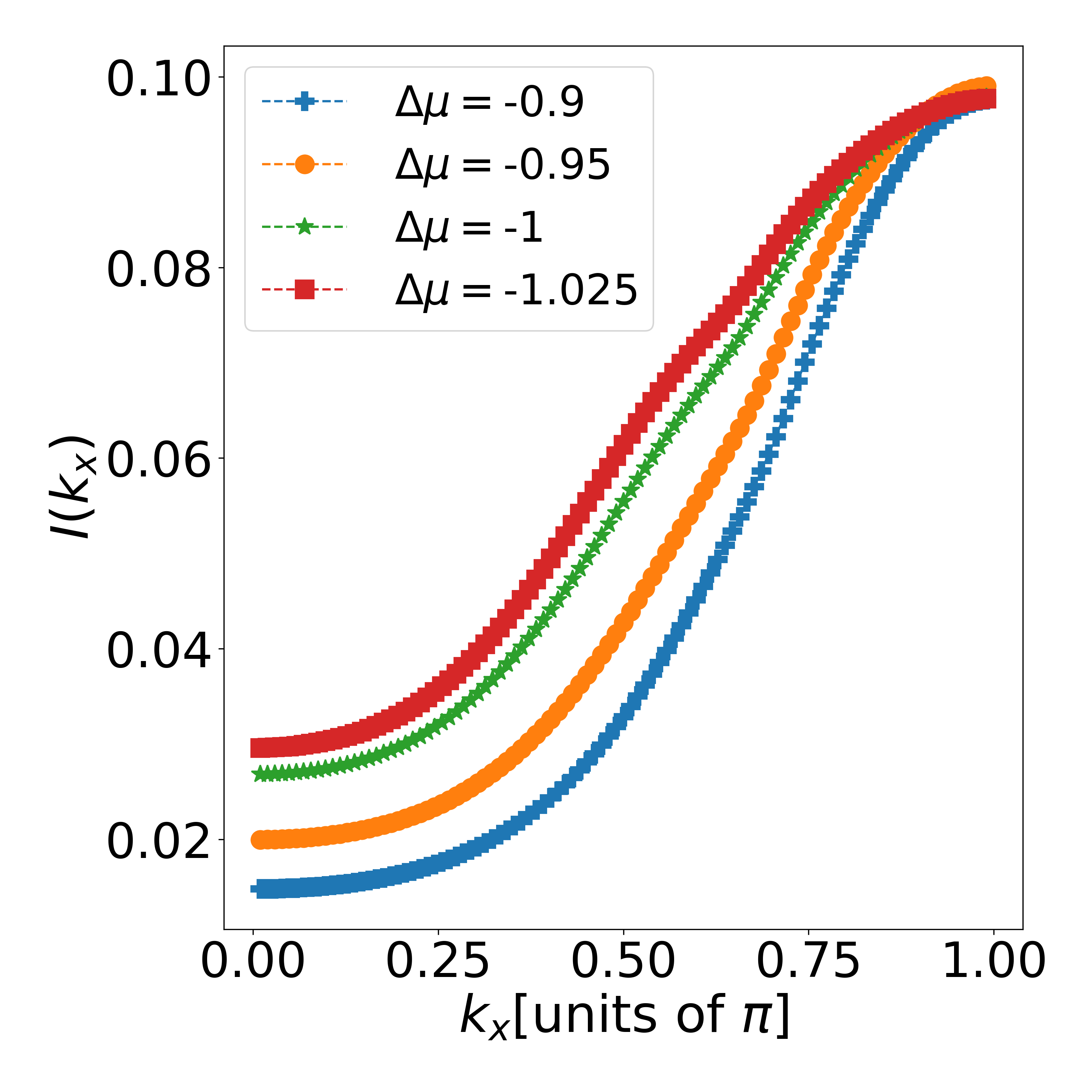}
            }
    \caption{Left panel: Discrete Fourier spectrum of $\mathcal{N}(\ell)$ for a slab with $N=101$ where all layers share the same hole-like Fermi surface. The spectra are very smooth and significantly less peaked than in the hole-like case. Right panel: $I(k_x)$ for the same values as in left panel. }
    \label{fig:FFT-e-integrale-hole-like} 
\end{figure}

\begin{figure}[ht]
    \centering
    \includegraphics[width=0.4\textwidth]{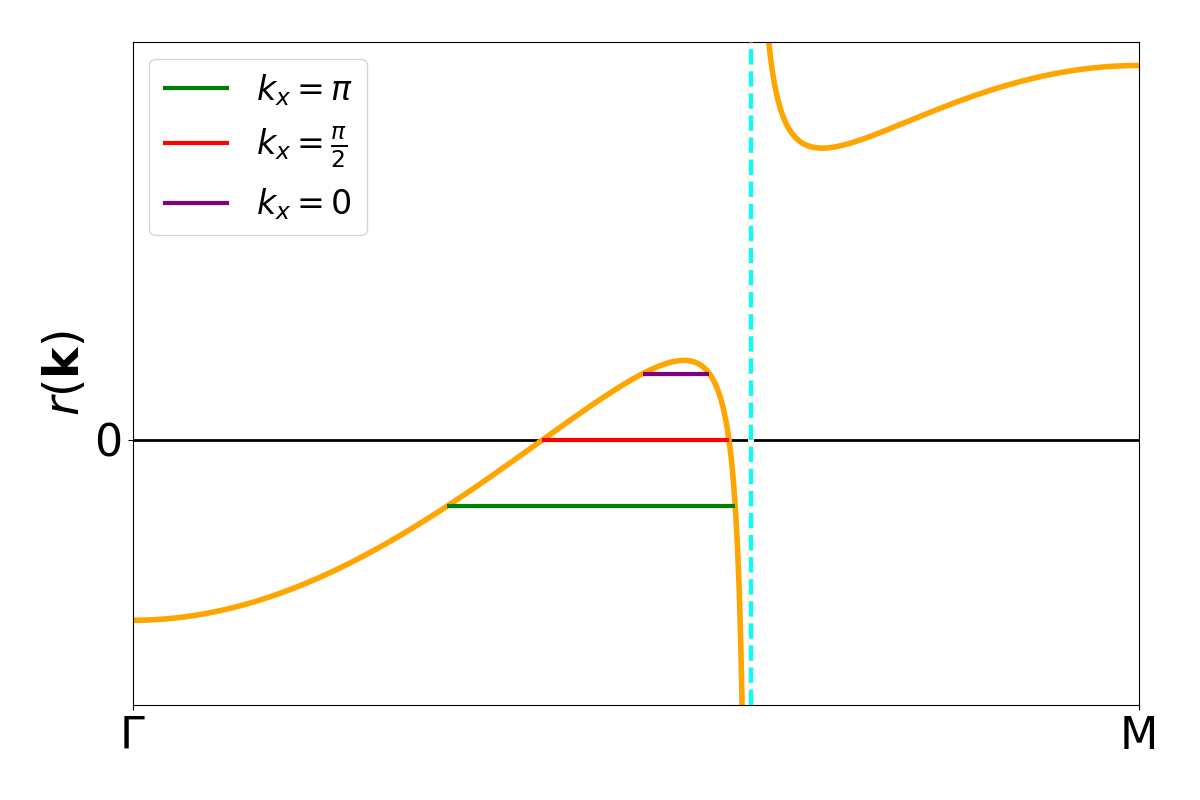}
            \caption{Sketch of the qualitative behavior of the renormalized dispersion $r(\bk)$ 
along $\mathbf{\Gamma} \bM$ in the case of Fermi pockets. The cyan line represents the Luttinger surface. The horizontal segments are the diameters of the Fermi pocket sheets along for a given $k_x$, showing that $k_x=\pi$ corresponds to the largest one.}
    \label{fig:sketch_pocket}
 \end{figure}

Finally, we investigate how the period of the FOs and $I(k_x)$ differ at low doping.    
In the left panel of Fig.~\ref{fig:FFT-e-integrale-hole-like} we plot the discrete Fourier transform $\mathcal{N}(K)$ of $\mathcal{N}(\ell)$ at the values of $\Delta \mu$ for which the Fermi surface is hole-like. Here, the scenario is completely different: the peak intensity decreases with doping, while its position increases.
In Fig.~\ref{fig:FFT-e-integrale-hole-like}, right panel, we instead plot $I(k_x)$ for the same 
$\Delta\mu$s, which now does not show a peak but rather a maximum for every value of $\Delta \mu$ at the zone boundary, $K_{M}=\pi$. Moreover, $I(k_x)$ decays slowly when $k_x$ is moved away from the maximum. This implies that the sum over $k_x$ in \eqn{eqn:I_con_sigma} cannot be estimated anymore by the saddle point approximation, which is also the reason of the smooth Fourier spectrum of $\mathcal{N}(\ell)$. This behavior of $I(k_x)$ at odds with the case of non-interacting electrons earlier discussed can be rationalized by the presence of Fermi pockets.\\
In the scenario proposed by Stanescu and Kotliar \cite{PhysRevB.74.125110} for the 2D weakly hole-doped Hubbard model, Fermi pockets coexist with a Luttinger surface whose position is determined by the line of poles of the zero frequency self-energy. 
In other words, the hole-like Fermi pockets appear around the maxima of the renormalized dispersion 
$r(\bk)$ that are caused by the pole singularity at the Luttinger surface, as illustrated 
in Fig.~\ref{fig:sketch_pocket} where we sketch the behavior of $r(\bk)$ along $\mathbf{\Gamma}\to \bM$. If we consider a slab composed of identical layers, each realizing the aforementioned scenario, the Fermi surface again consists of $k_x$-dependent sheets defined by
\be
r\big(\bk_F(k_x)\big) =  2t\,\cos k_x\,.
\ee
It readily follows that the largest Fermi surface sheet is obtained at $k_x=\pi$, see  Fig.~\ref{fig:sketch_pocket}. This conclusion holds irrespective of $\mu$ as long as the 2D Fermi surface features hole-like Fermi pockets. Furthermore, since the largest Fermi surface sheet at 
$k_x=\pi$ is not an extreme point, unlike in the non-interacting case, we also understand why 
$I(k_x)$ is a smooth function of $k_x$.

\begin{figure}[H]
    \centerline{
            \includegraphics[width=0.23\textwidth]{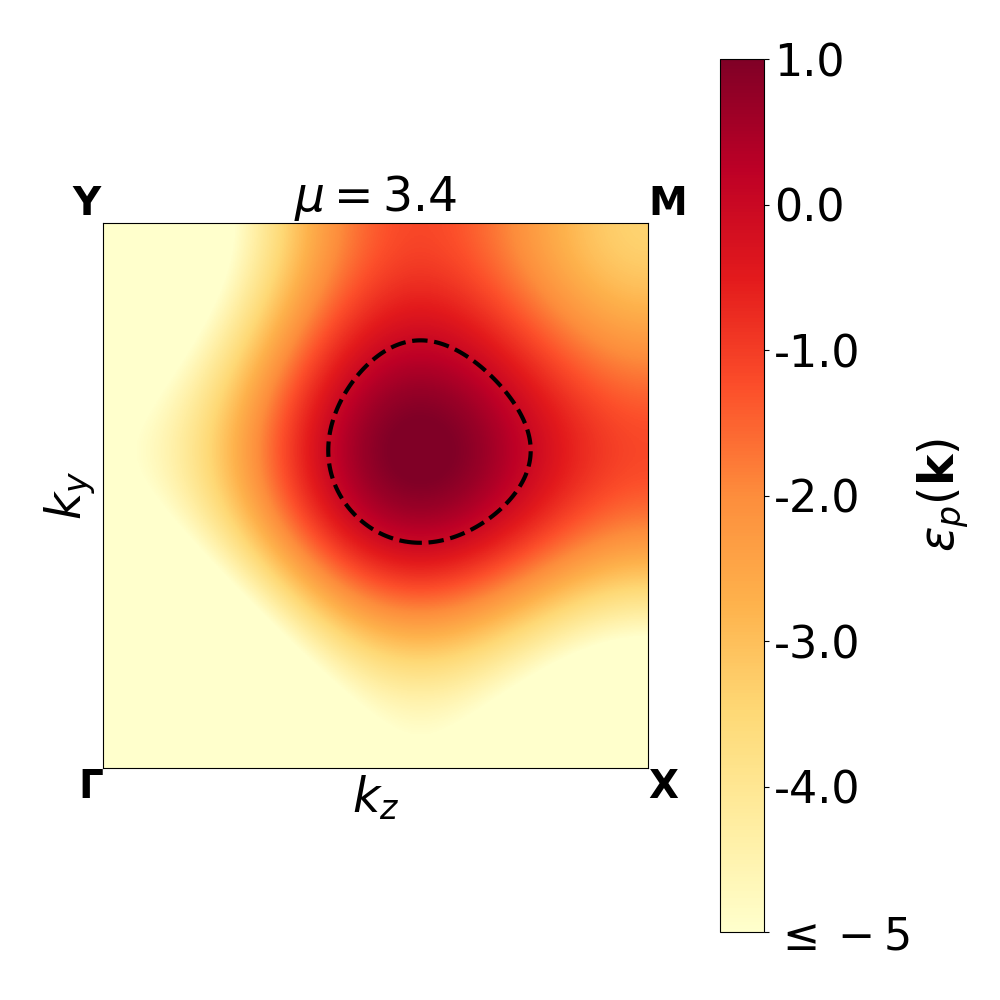}
            \includegraphics[width=0.23\textwidth]{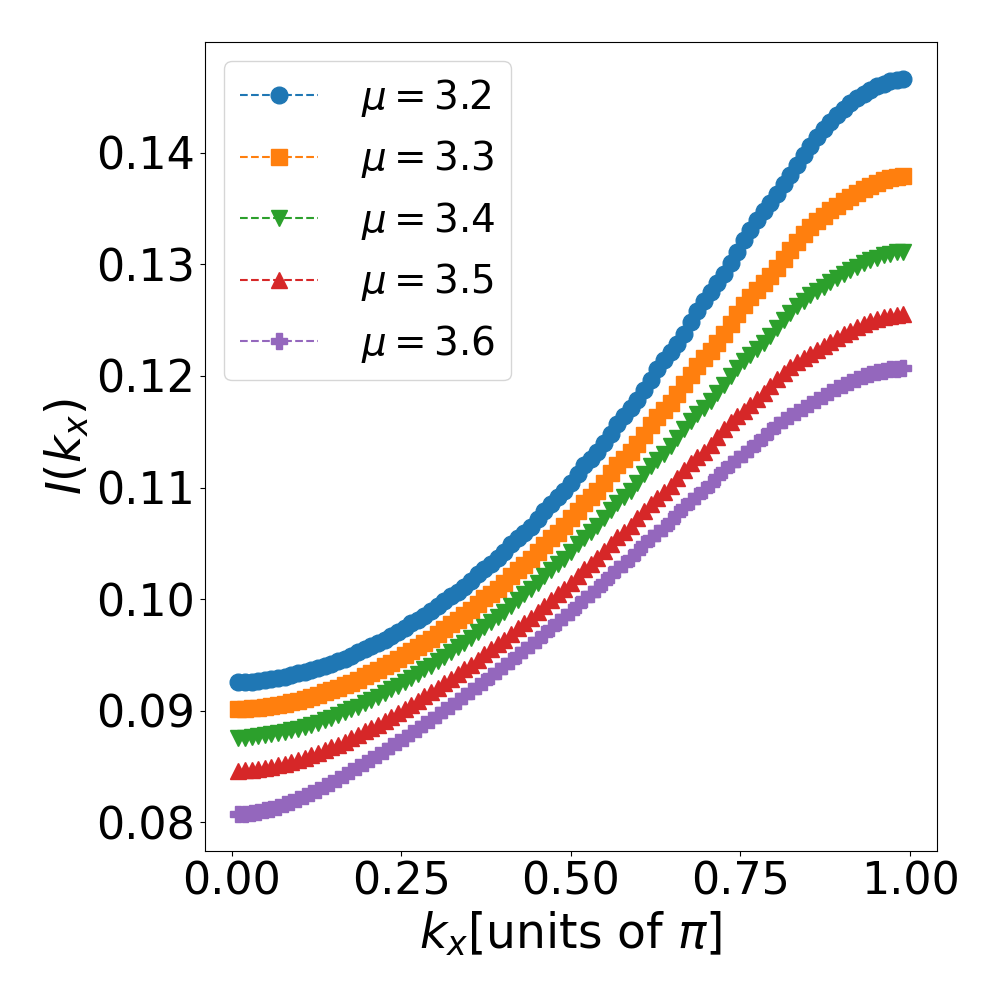}
            }
    \caption{Left panel: Color map of the in plane dispersion $\ep_p(\mathbf{k})$ in
    the upper right quarter of the Brillouin zone for the hopping values reported in text and $\mu=3.4$. The dashed line corresponds to the Fermi surface, which consist of a closed hole pocket. Right panel: $I_0(k_x)$ computed using Eq.~\eqn{eqn:ep} as in-plane dispersion for different values of $\mu$. As in Fig.~\ref{fig:FFT-e-integrale-hole-like}, right panel,  the maximum is always located at $K_M=\pi$.}
    \label{fig:pockets-artificiali} 
\end{figure}

\noindent
To better understand the link between the position of the maximum of $I(k_x)$ and the Fermi pockets, we study a simple non-interacting 2D tight-binding model that possesses Fermi pockets, whose 
energy dispersion in momentum space reads 
\beal
\ep_p(\bk) &=-2t\,\big(\cos k_z+\cos k_y \big)-4t'\,\cos k_z\, \cos k_y \\
&\qquad  -2t''\,\big(\cos 2 k_z+\cos 2 k_y\big) -\mu \;.
\label{eqn:ep}
\eal
We take $t=t''=1$ and $t'=0$, in which case for $\mu > 4.5$ the band is completely filled, while, for $\mu< 4.5$, a Fermi surface appears and it is made of hole pockets, see the left panel of Fig.~\ref{fig:pockets-artificiali}. The pockets persist until $\mu\simeq 2.2$ where the system undergoes a Lifshitz transition to a closed Fermi surface. Now, we consider the slab geometry,   
with dispersion $\ep_p(\bk)-2t_\perp\,\cos k_x$ and number of layers $N=101$, and compute $I_{0} (k_x) $. We take $t_{\perp}=0.33$ to reproduce a setting in which all Fermi surface sheets at different $k_x$ are still made of closed hole-like Fermi pockets. We plot $I_{0} (k_x) $ in Fig.~\ref{fig:pockets-artificiali}, confirming that its maximum is always at $k_x=\pi$ for every value of $\mu$ with hole pockets. Moreover, the behavior of $I_0(k_x)$ as $\mu$ decreases is qualitatively similar to the one of the interacting $I(k_x)$ at weak hole-doping. \\
It is also interesting what happens if one computes $I_{0} (k_x)$ considering a Fermi surface that is hole-like but does not feature Fermi pockets.  This setting can be realized by taking $t=1$, $t'=-0.3 $ and $t''=0$ in \eqn{eqn:ep}, parameters often used to model the non-interacting dispersion of high-$T_c$ superconductors \cite{tprimo}. In this case, $K_M$ is not equal to $\pi$ and depends instead on the chemical potential.
Therefore, the condition $K_M=\pi$ for the maximum of $I(k_x)$ is not simply a marker of the hole-like character of the Fermi surface, but rather of hole-like Fermi pockets. \\
In conclusion, $I(k_x)$ can identify fingerprints of Fermi pockets from a simple 4-patches DCA calculation without relying on any cluster periodization scheme, thus complementing the information obtained by the self-energy and the Green's function.     

\section{Conclusions} \label{conc}
We have demonstrated that the surface dead layer effect \cite{dead-layer} can lead to a remarkable phenomenon: the reconstruction of the Fermi surface as a function of the distance from the surface in a hole-doped Mott insulator. This effect occurs because the more correlated surface layers are less doped, and, as evidenced by several numerical studies in single layers, they host hole-like Fermi pockets that violate Luttinger's theorem. In contrast, the doping increases in the less correlated inner layers, which display the expected electron-like Fermi surface.We have also introduced a novel quantity, $S(T)$ in \eqn{ST}, which serves to directly diagnose the breakdown of Luttinger's theorem.  
\\
Furthermore, we have investigated how the different topology of the Fermi surface at weak and large hole-doping affects the Friedel oscillations in a slab geometry. When the doping is weak, the Friedel oscillations are shown to be consistent with the existence of hole-like Fermi pockets. Conversely, at higher doping, the Friedel oscillations bear the hallmarks of an electron-like Fermi surface adiabatically connected to the non-interacting one.\\
Both the Friedel oscillations and $S(T)$ in \eqn{ST} are therefore effective tools to 
characterize the Fermi surface topology, without the need of any periodization scheme that may introduce biases when extracting the Fermi surface from the Green's function. \\
 The layer-dependent Fermi surface reconstruction and Friedel oscillations are phenomena unique to slab geometry that are absent with periodic boundary conditions. Our findings demonstrate how open boundary conditions introduce additional features that could be crucial in interpreting surface-sensitive spectroscopic data. 

\noindent
\textit{Note Added: } After completing this work, a preprint titled "Probing Quasiparticle Excitations in Doped Mott Insulators" \cite{banerjee2025probingquasiparticleexcitationsdoped} has been posted on the arXiv. This preprint investigates Friedel oscillations induced by an impurity in a weakly doped Mott insulator using a different technique. The authors draw the same conclusions as ours, namely, that Friedel oscillations exhibit distinctive properties that reflect the breakdown of Luttinger's theorem.

\section*{Acknowledgments}
We acknowledges useful discussions with Antonio Maria Tagliente, Ivan Pasqua and Erik Linn\'er. We acknowledge the CINECA award under the ISCRA initiative for the availability of high-performance computing resources and support.\\
 
\section*{appendix}
\subsection{Criterion for the identification of the pseudogap and relationship with the Fermi surface topology}

\begin{figure}[ht]
    \centering
    \includegraphics[width=0.45\textwidth]{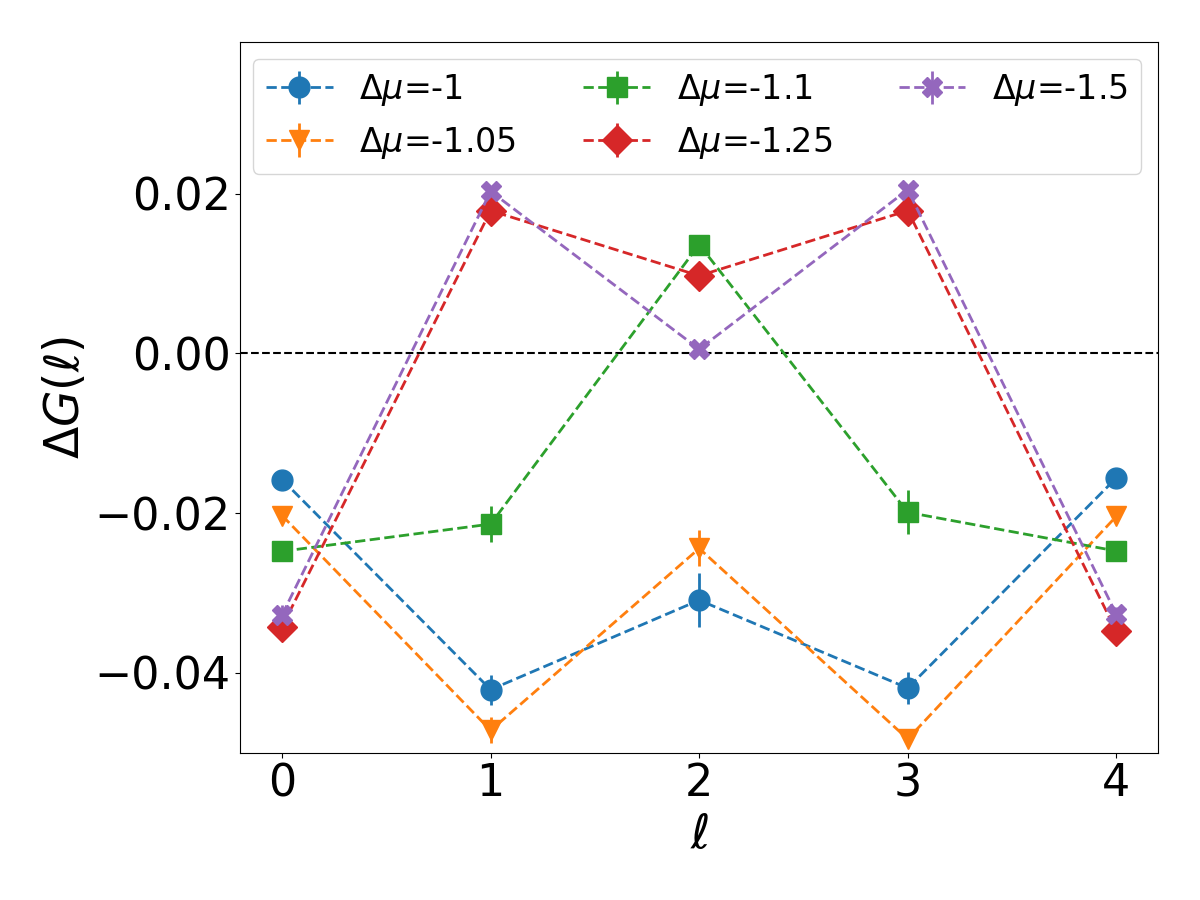}
    \caption{ $\Delta G(\ell)$ as a function of layer. Comparing it with Fig~.\ref{fig:reg_5L and fig:n5_L}, left panel, we observe that, whenever the layer has a hole-like Fermi surface, a pseudogap is also present. }  
    \label{fig:pg_criterion}
\end{figure}

Following Ref.~\cite{Georges-PRX2018}, we determine if a layer $\ell$ is in the pseudogap state by analyzing the behavior of the imaginary part of the Green function at zero frequency within patch $\mathbf{X}$.  This quantity is obtained through a linear extrapolation at the first two fermionic Matsubara frequencies. In a pseudogap metal,  $-\Ima\, G (i\omega \rightarrow 0, \mathbf{X},\ell) $ decreases with temperature, while in a correlated Fermi liquid the scenario is opposite. Therefore, we introduce the quantity
\beal
 \Delta G(\ell) &=-\fract{1}{\pi}\,\Big(\Ima\, G (i\omega \rightarrow 0, \mathbf{X},\ell)_{\big\lvert \beta=40}  \\
&\qquad\qquad\;   -  \Ima G (i\omega \rightarrow 0, \mathbf{X},\ell)_{\big\lvert \beta=20} \Big) \,,
\eal  
with $\beta=T^{-1}$. The layer $\ell$ is considered pseudo gapped if $\Delta G(\ell) < 0$. We plot $\Delta G(\ell)$ in Fig.~\ref{fig:pg_criterion}. Comparing this figure with Fig.~\ref{fig:reg_5L and fig:n5_L}, left panel, we observe that a pseudogap is present whenever the layer has a hole-like Fermi surface, in good agreement with previous results \cite{Georges-PRX2018,PhysRevLett.120.067002}. 

\subsection{Fermi liquid behavior of the self-energy}

\begin{figure}[ht]
    \centering
    \includegraphics[width=0.45\textwidth]{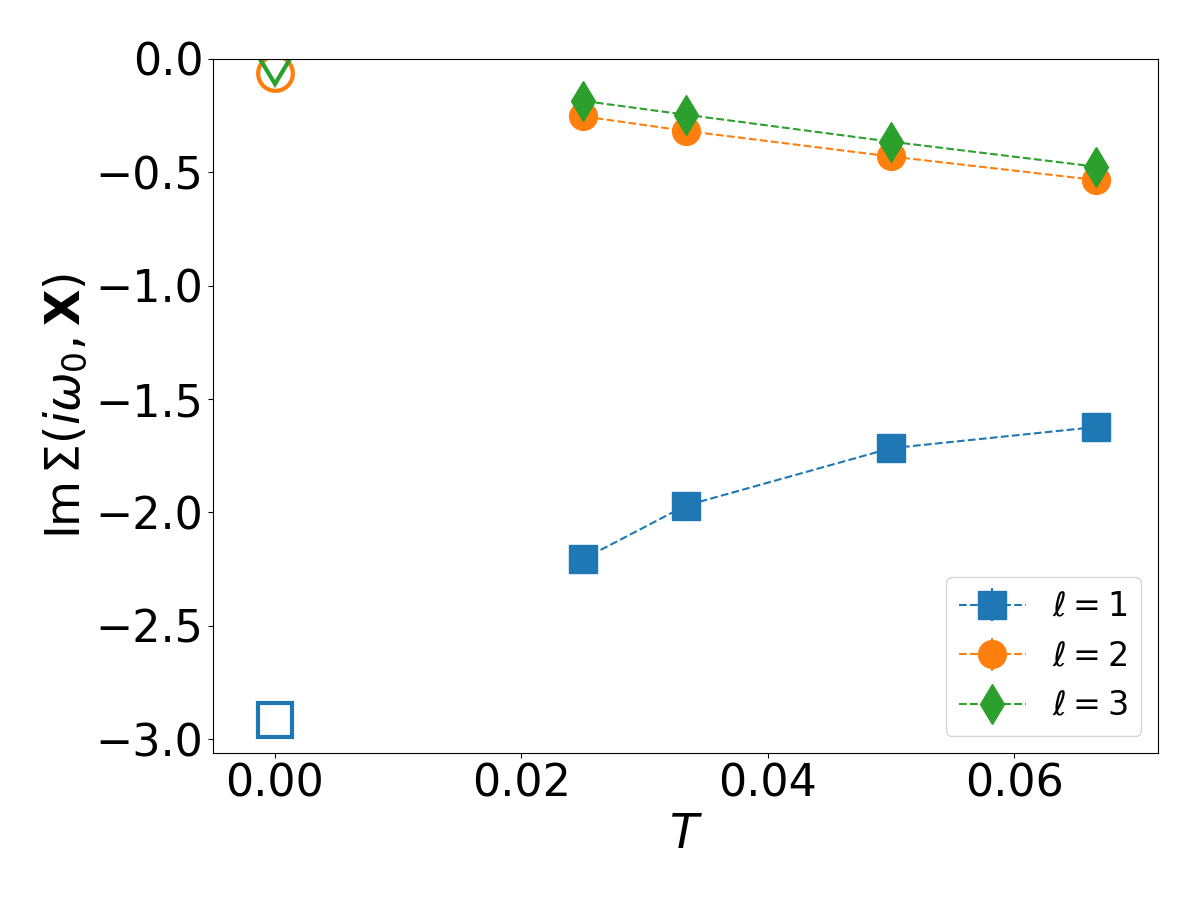}
    \caption{  $ \Ima \Sigma_{\ell}( i \omega_0,\mathbf{X})$ as function of $T$ for the first three layers of the slab with $\Delta \mu=-1.5$. The points with empty marker represent the $T\to 0$ value of the self-energy, obtained with a linear extrapolation of the two lowest values of $T$.   }
    \label{fig:FL}
\end{figure}

The dead-layer phenomenon can lead to the coexistence of Fermi surfaces with different topologies within the same slab, with the outer layers displaying a hole-like Fermi surface and the inner layers  an electron-like one. To better characterize the two cases, we examine the temperature dependence of the imaginary part of the self-energy at the first Matsubara frequency. In a conventional Fermi liquid, this quantity begins linear in $T$ \cite{First-Matsubara}. We focus on patch $\mathbf{X}$, where 
correlation effects are stronger \cite{Phase_diagram_DCA4}.
In Fig.~\ref{fig:FL} we plot $\Ima\, \Sigma_{\ell}(i\omega_0,\mathbf{X})$ as a function of $T$ for $\Delta\mu=-1.5$. For this chemical potential value, the second and third layers exhibit an electron-like Fermi surface, and their imaginary part of the self-energy displays a linear dependence on temperature, with an extrapolated $T=0$ value approaching zero, indicating conventional Fermi-liquid behavior. In contrast, the first layer, which has a hole-like Fermi surface, exhibits a clear nonlinear temperature dependence in its self-energy, and the extrapolated $T\to 0$ value remains finite and large. However, we emphasize that $\Ima\, \Sigma_{\ell}(i\omega_0\to 0,\mathbf{X})\not= 0$, which 
corresponds to a vanishing quasiparticle residue,  
does not necessarily imply non-Fermi liquid behavior, as discussed in \cite{Michele2}. 
In fact, coherent quasiparticles exist if their decay rate
decays at least as $\ep^2$, where $\ep$ is the real frequency. This decay rate is, 
rigorously speaking, 
\be
\gamma(\ep,\bk) = -Z(\ep,\bk)\,\Ima\,\Sigma(\ep,\bk)\,,
\label{qp-gamma}
\ee
where $Z(\ep,\bk)$ is the quasiparticle residue and $\Sigma(\ep,\bk)$ the retarded self-energy, i.e., the analytic continuation of $\Sigma(i\omega_n,\bk)$ for $i\omega_n\to \ep + i0^+$. 
$\gamma(\ep,\bk)$ in \eqn{qp-gamma} vanishes as $\ep^2$, thus exhibiting Fermi liquid-like behavior, 
if $ -\Ima\,\Sigma(\ep,\bk)\simeq \ep^2$ and $Z(\ep\to 0,\bk)>0$, as in conventional Fermi liquids, 
but also if $ -\Ima\,\Sigma(\ep\to 0,\bk)>0$ and $Z(\ep,\bk)\simeq \ep^2$, which, despite the apparent contradiction, does not invalidate Fermi liquid theory \cite{Michele2}.

\subsection{ DCA self-energy in the 5-layer slab}
\begin{figure*}[h]
    \centering

    \begin{subfigure}[t]{0.6\textwidth}
        \centering
        \includegraphics[width=\textwidth]{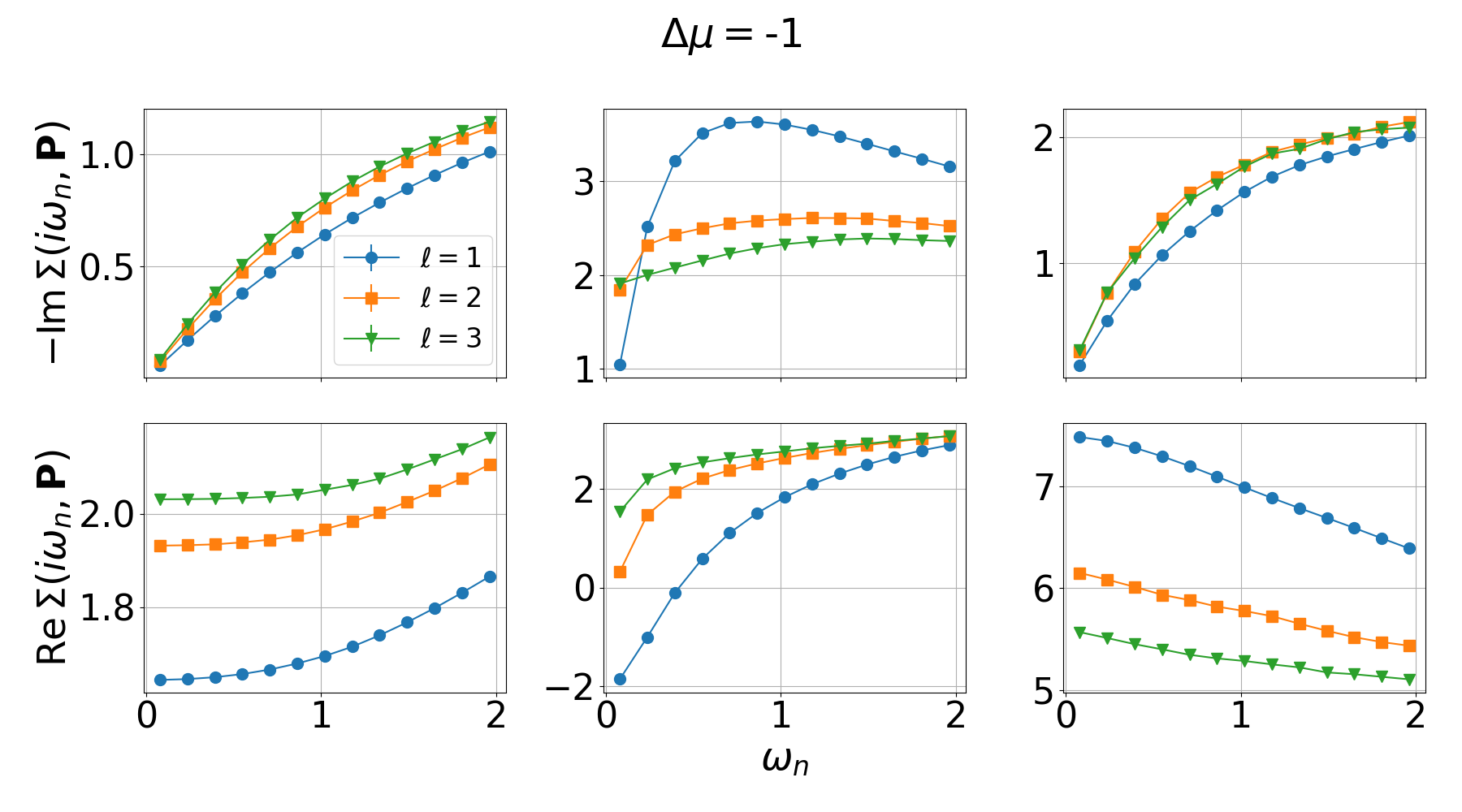}
        
    \end{subfigure}
    
    \vspace{0.5em}  

    \begin{subfigure}[t]{0.6\textwidth}
        \centering
        \includegraphics[width=\textwidth]{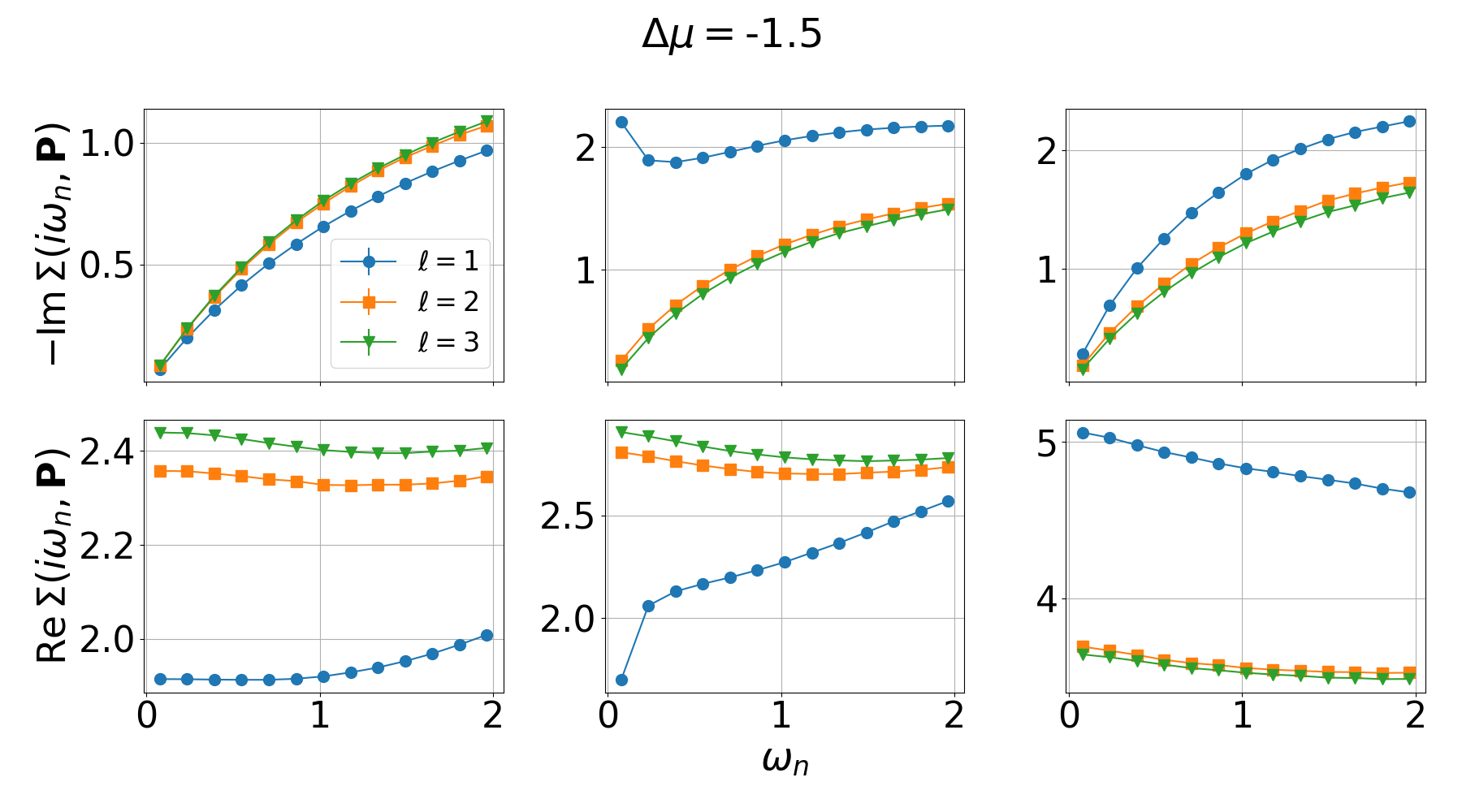}
        
    \end{subfigure}

    \vspace{0.5em}

    \begin{subfigure}[t]{0.6\textwidth}
        \centering
        \includegraphics[width=\textwidth]{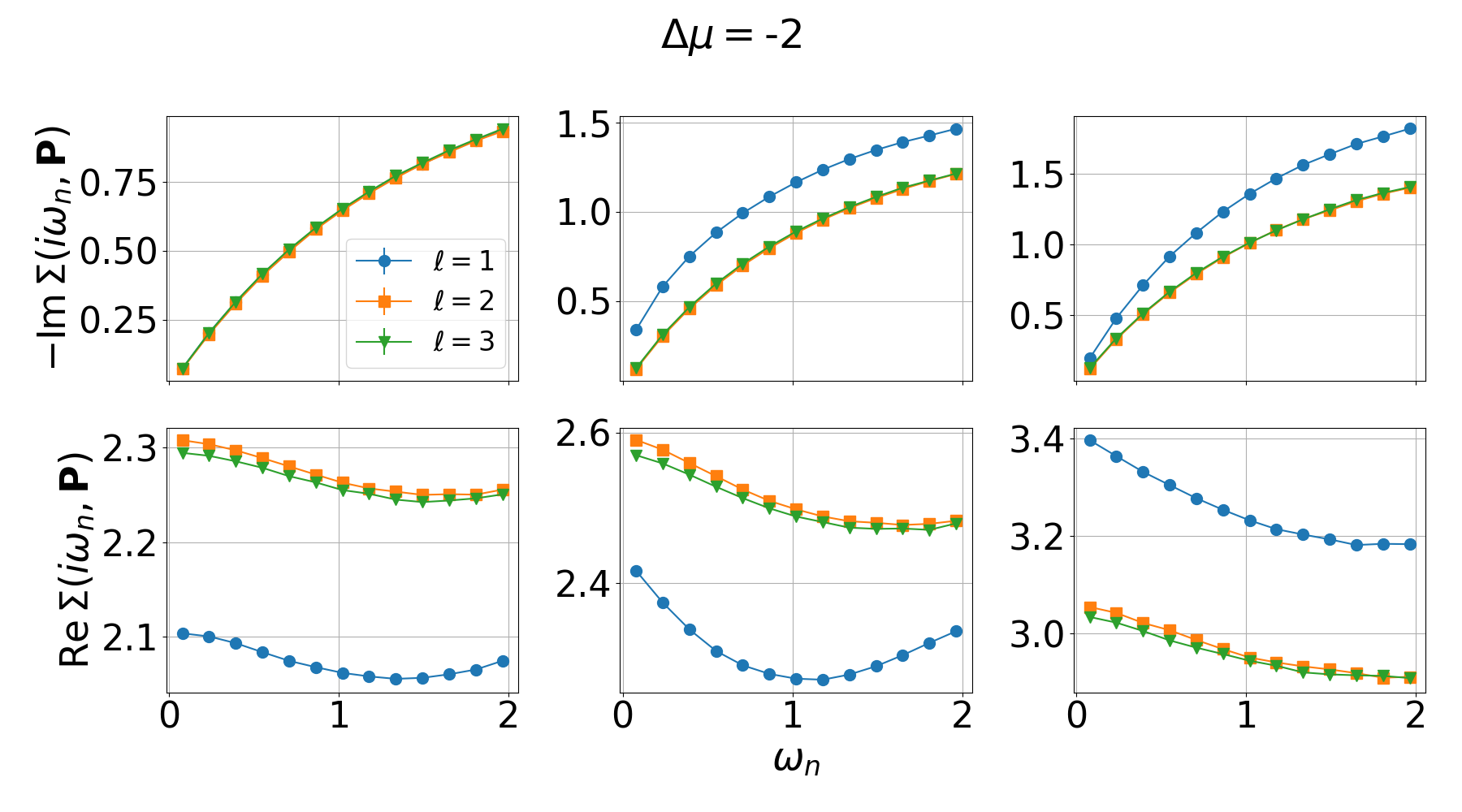}
    \end{subfigure}

    \caption{ Imaginary and real part of the self energy as function of $\omega_n$ for the first three layer of the slab. Left column refers to patch $\mathbf{\Gamma}$, central column to $\mathbf{X}$ and right column to $\bM$. The values of the chemical potential correspond to a slab where all the layers have an hole-like Fermi surface ($\Delta \mu=-1$), to a slab where only the edge layers are hole-like ($\Delta \mu =-1.5)$ and, finally, to a slab where all layers are electron-like ($\Delta \mu=-2$).}
    \label{fig:self_energies}
\end{figure*}

{ In Fig.~\ref{fig:self_energies} we plot the real and imaginary part of the DCA self-energies in Matsubara frequencies for three different patches and values of $\Delta \mu$. We observe that, for every value of the chemical potential,  the imaginary part of the self-energy in patch $\mathbf{X}$ is significantly more layer-sensitive if compared with patch $\bM$ and $\mathbf{\Gamma}$. } 

\FloatBarrier

\end{document}